# Drag reduction study of naturally occurring oscillating axial flow induced by helical corrugated surface in Taylor Couette flow

Md Abdur Razzak[1], Khoo Boo Cheong (邱武昌)[1], Kim Boon Lua (賴錦文)[2] and C.M. J. Tay[1]

[1] Department of Mechanical Engineering, National University of Singapore, Singapore 119260, Singapore

[2] Department of Mechanical Engineering, National Chiao Tung University, Taiwan

**Abstract:** This study investigates drag reduction capability of naturally-occurring-oscillating axial secondary flow(ASF) induced by helical-corrugated surface in Taylor Couette flow(TCF$_{Helical}$) for three values of pitch to wavelength-ratios($P*$ =1,2,3) and amplitude to wavelength-ratio($A*$) of 0.25. As reported in Razzak et al. (2020), emergence of naturally-occurring-oscillating ASF induced by longitudinal-corrugated surface in TCF(TCF$_{Longitudinal}$) and increasing trend on its magnitude with Reynolds number ($Re$) results in the occurrence of drag reduction. This has motivated us to study the possibility of enhancing drag reduction by maintaining a consistently increasing trend with $Re$ in the magnitude of naturally-occurring-oscillating ASF induced by the helical-corrugated surface on the stationary outer cylinder in TCF. From flow structures, steady ASF with non-zero mean is observed at $Re$=60 which suppresses the strength of azimuthal vorticities for $Re$>85 and contributed to occurrence of drag reduction. As $Re$ is increased to 100,90 and 85 for $P*$ =1,2, and 3, respectively, formation of periodic oscillating ASF with non-zero mean and its increasing trend in magnitude with $Re$ suppresses azimuthal vorticities further which contributes to the maximum drag reduction of 13%. For $Re$>165,145 and 140 for $P*$=1,2 and 3, respectively, non-periodic oscillating ASF is observed, and its magnitude remains nearly unchanged or decreases slightly with $Re$ which results in the suppression effect of azimuthal vortices to be weaker. This results in decrease in drag reduction. Oscillating ASF observed in TCF$_{Helical}$ is found to occur at earlier $Re$ and it is stronger than that of TCF$_{Longitudinal}$ which contributes to occurrence of higher drag reduction in TCF$_{Helical}$.

## 1 Introduction

Global warming is one of the greatest threats to existence of life on earth and is mainly due to the carbon emissions from burning fossil fuel. The emission of billions of tons of $CO_2$ is directly responsible for global warming and frequent natural disasters. Aviation and shipping consume of a major portion of world's total fossil fuels and contribute a large amount of carbon emissions yearly. Even so, the demand for fuel consumed by these sectors is increasing every year and this translates into expenditures or billions of dollars to meet the fuel requirements for these sectors annually. Therefore, even a small percentage reduction in fuel consumption can save billions of dollars and a reduction in a huge amount of carbon emission yearly. The potential economic gains and the added benefit of reducing its carbon footprint have led researchers to search for effective means of reducing fuel consumption over the last century.

Skin friction drag is considered as one of the dominant sources of energy expenditure in aircraft (45%), ships (50%), submarines (60%), and piping systems (100%) (Perlin et al., 2016). Thus, reducing skin friction drag is one of the most effective ways of enhancing energy efficiency. The potential economic benefit has led researchers to devote their efforts to find efficient means and strategy to reduce skin friction drag by development of several drag reduction techniques over last few decades. Past studies involved the use of polymers, riblets, dimples, surfactants, superhydrophobic surfaces, air bubbles, wall oscillation and laminar flow control ( Ferrante & Elghobashi, 2004; Fish & Lauder, 2005; Walsh, 2008; García et a., 2011; Quadrio, 2011; Vakarelski et al., 2011; Perlin et al., 2016; Bhambri & Fleck, 2016). However, the drag reduction achieved by application of the above-mentioned techniques is limited along with additional complexities involved in manufacturing and maintenance. With a view to overcoming the limitations of previously developed techniques, a more effective way of reducing drag reduction was reported with the use of macroscale longitudinal corrugated surface (Mohammadi & Floryan, 2012, 2013a, 2013b, 2015, 2019; Chen at el., 2016; DeGroot et al., 2016; Ghebali et al., 2017; Ng et al., 2018;

Yadav et al., 2018; Moradi & Floryan, 2019) where "macroscale" refers to surface modifications with a length scale comparable to that of the channel opening (Mohammadi & Floryan, 2012, 2013a, 2013b; Chen et al., 2016; DeGroot et al., 2016). One may consider corrugated surfaces as riblets. However, there are distinct differences between riblets and corrugated surfaces. The most fundamental differences are in their characteristics size. Riblets are microscale surface roughness where wavelength of riblets is the order of the viscous scale while wavelength of macroscale corrugated surface is in order of channel height (Mohammadi & Floryan, 2012, 2013a, 2013b; Chen at el., 2016; DeGroot et al., 2016). This translates dimensions of characteristics size of corrugated surfaces can be up to 1000 times larger than riblets due to which manufacturing, and maintenance of corrugated surfaces are easier than that of riblets which suggests use of macroscale corrugated surface (i.e., corrugated surface) in drag reduction to be a more effective and sustainable techniques.

A total of nearly 5% drag reduction was reported for both laminar channel flow (Mohammadi & Floryan, 2013) and turbulent channel flow (Chen et al., 2016) using longitudinal corrugated surface. Yadav et al. (2018) had studied a rectangular duct with a corrugated top-bottom wall to find the characteristics of geometry for early flow destabilization which may be useful in drag reduction studies. In a recent study, Moradi & Floryan (2019) had investigated the flow instability involved in annuli with longitudinal grooves. Ghebali et al.(2017) studied oblique corrugated surface in turbulent channel flow and reported reduction in skin friction drag. Though the above-mentioned studies had been conducted with macroscale corrugated walls, the mechanism involved in the drag reduction using corrugated surfaces is not well understood yet. Thus, the main focus of this study is to investigate the drag reduction capability of macroscale corrugated surfaces and the detailed overview of its possible drag reduction mechanism.

Experimental and numerical investigations of skin friction drag reduction usually involve facilities such as open turbulent boundary layer flow, channel flow or pipe flow. However, the limitations and complexities of measuring small frictional forces in experiments and the huge computational costs involved in numerical simulation of these facilities have led to the use of Taylor-Couette flow (TCF) configuration in recent years as an alternative means of studying drag reduction. The attraction of TCF configuration is partly due to its compactness and ease of measuring torque or friction forces with a higher degree of accuracy. This has led to numerous investigations of drag reduction since the 1970s using the TCF configuration (Quan, 1972; Groisman & Steinberg, 1996; Yi & Kim, 1997; Kalashnikov, 1998; Watanabe & Akino, 1999; Hall & Joseph, 2000; Koeltzsch et al., 2003; Murai et al., 2005; Climent et al., 2007;  Sugiyama et al., 2008; Dutcher & Muller, 2009; Greidanus et al., 2011; Eskinn, 2014; Srinivasan, et al., 2014; Greidanus et al., 2015; Bhambri & Fleck, 2016; Gao et al., 2016; Zhu et al., 2016; Rosenberg et al., 2016 ). The success of past studies motivates the use of the TCF configuration in the present drag reduction study with a corrugated surface.

The motion of viscous fluid in between two coaxial rotating cylinders, also known as the TCF configuration usually involves two coaxial cylinders with either one fixed cylinder and another rotating, or both cylinders rotating in either the same or opposite directions. From previous studies, TCF has been used to study drag reduction using polymers, riblets, surfactants, superhydrophobic surfaces and air bubbles (Quan, 1972; Groisman & Steinberg, 1996; Yi & Kim, 1997; Kalashnikov, 1998; Watanabe & Akino, 1999; Hall & Joseph, 2000; Koeltzsch et al., 2003; Murai et al., 2005; Climent et al., 2007; Gao et al., 2016; Zhu et al., 2016). Although not directly related to the drag reduction, based on the stability analysis, Floryan (2002) reported the influence of centrifugal effects on the formation of streamwise vortices using a stationary transversely corrugated outer cylinder and rotating smooth inner cylinder in TCF. Ikeada & Maxworthy (1994) investigated instability in TCF using a rotating inner cylinder with a sinusoidal varying groove along the axial direction and stationary outer cylinder. The purpose was to investigate the instability under the application of external force originating from the corrugated surface in TCF. It was observed that the corrugated surface affects only the size of the Taylor vortex and not the onset of Taylor instability. To the best of our knowledge, Ng et al.( 2018) studied a corrugated surface for the first time with a stationary outer cylinder and a smooth rotating inner cylinder for the radius ratio of 0.5 and at Reynolds numbers ($Re$) 20 and 100. The drag reduction obtained in their study was then compared with the results of Fasel & Booz (1984) who conducted a similar study for the same radius ratio 0.5 but with both smooth inner and outer cylinders.  This study had reported higher drag with the application of corrugated surface.

For the same radius ratio used by Ng et al. (2018), Razzak et al. (2019) observed that the emergence of naturally occurring periodic oscillating axial flow in TCF with a smooth cylinder(TCF$_{Smooth}$) is found to suppress the strength of azimuthal vortices (i.e., streamwise vortices) for $Re > 425$ which in turn contributes to the reduction in wall shear stress. The behaviour of naturally occurring periodic oscillating axial secondary flow reported in Razzak et al. (2019) is very similar to the application of spanwise wall oscillation for drag reduction reported in many studies eg. Mangiavacchi et al.(1992), Laadhari et al.(1994), Baron & Quadrio (1996), Choi & Graham(1998), Choi & Clayton (2001), Skote (2011;2012), Hehner et al.(2019) and Yao et al.(2019), where spanwise wall oscillation suppresses the strength of the streamwise vortices (i.e., streamwise vorticity) and pushes the vortices away from wall, leading to a reduction in the wall shear stress with increasing $Re$. Interestingly, the oscillating axial secondary flow reported in Razzak et al.(2019) occurs naturally and does not need any external power or complex system otherwise used in the application of spanwise wall oscillation for drag reduction reported in the above mentioned studies. Therefore, the use of surface manipulation to trigger and enhance the emergence of naturally occurring oscillating axial flow would be a more effective, sustainable and practical drag reduction technique.

Based on this, Razzak et al. (2020) performed experimental and numerical investigation of TCF with a longitudinally corrugated stationary outer cylinder(TCF$_{\text{Longitudinal}}$) and a rotating smooth inner cylinder for $Re$ between 60 and 650, keeping the radius ratio the same as that of Razzak et al. (2019) to study the behaviour of naturally occurring oscillating axial secondary flow and its corresponding contribution to the wall shear stress under the influence of a corrugated surface in detail. It was found that the application of longitudinal corrugated surfaces resulted in the emergence of naturally occurring oscillating axial secondary flow at an earlier $Re$ (i.e., $Re$ =125) and contributed to a maximum of 6.08% drag reduction. In addition, drag reduction in TCF with longitudinal corrugated surface was found to increase when the magnitude of the oscillating axial secondary flow increases and vice-versa. This suggests that if it is possible to maintain a consistently increasing trend in the magnitude of oscillating axial secondary flow with $Re$ (i.e., avoiding the decreasing phase), this may help to (continuously) enhance the drag reduction. This has motivated us to study the generation of a stronger oscillating axial secondary flow induced by longitudinal corrugated surface arranged in a helical path (i.e., helical corrugated surface) on a stationary outer cylinder and a rotating smooth inner cylinder in TCF and its contribution on drag reduction. It can be noted that the study of TCF with a helical corrugated surface (TCF$_{\text{Helical}}$) has not been reported in the earlier studies. In the present study, we would like to better understand the behaviour of naturally occurring oscillating axial secondary flow and its corresponding contribution to drag reduction under the influence of a helical corrugated surface. We would also like to better understand the flow structures and their variation with $Re$ under the influence of a helical corrugated surface. In the present study, a TCF of radius ratio = 0.5 has been studied for three pitch to wavelength ratios ($P^*$) of 1, 2 and 3, amplitude to wavelength ratios ($A^*$) of 0.25, amplitude to gap ratio of 0.5 and $Re$ ranging between 60 and 650.

Because of the very large computational resources required by Direct Numerical Simulation (DNS) schemes, Large Eddy Simulation (LES) was employed as an alternative numerical tool to study TCF in previous studies (Bazilevs & Akkerman , 2010; Salhi et al., 2012; Bauer et al., 2013; Poncet et al., 2013; 2014; Wang et al., 2016; Ohsawa et al., 2016 and Razzak et al., 2020). This has motivated us to employ LES in the present study to investigate TCF with a helical corrugated surface. The numerical configuration used in the present study is exactly the same as that of Razzak et al. (2020) where LES was implemented for TCF with a longitudinal corrugated surface (TCF$_{\text{Longitudinal}}$). As described below, the torque was found to be independent of the height of cylinder for $H \geq 6d$ (i.e., $d$ = annular gap) which is in alignment with the finding of Razzak et al. (2020), where the flow structures and torque become independent of the height of cylinder for $H \geq 8d$. Therefore, a nominal height of cylinder $H= 8d$ has been selected with the periodic boundary condition at both ends of cylinders. Although the LES scheme used by Razzak et al. (2020) had already been extensively validated against DNS and experiments for TCF with longitudinal corrugated surfaces(TCF$_{\text{Longitudinal}}$), further validation of the LES used in the present study for TCF with a helical corrugated surface(TCF$_{\text{Helical}}$) was also conducted by comparing the LES results with DNS simulations of the same flow.

## 2 Configuration of helical-corrugated surface in TCF(TCF$_{\text{Helical}}$)

This study involves a sinusoidal corrugated surface being distributed along the axial direction and arranged in a helical path on the outer cylinder and a smooth inner cylinder as shown in *Figure 1a* and *Figure 1b* where the z-axis coincides with the central axis of the cylinders. The amplitude and wavelength of the sinusoidal corrugated surface are represented by $A$ and $\lambda$, respectively. The pitch of the helical path is denoted by $P$. The radius of the inner cylinder is $R_i$ and the mean radius of the outer cylinder is $R_m^o$. The computational height of cylinders is $H$ and the mean annulus gap is $d$ (i.e., $d = R_m^o - R_i$). The non-dimensional parameters involved in the TCF$_\text{Helical}$ are the amplitude to wavelength ratio ($A^* = \frac{A}{\lambda}$), the amplitude to gap ratio ($D^* = \frac{A}{d}$), the pitch to wavelength ratio ($P^* = \frac{P}{\lambda}$), the radius ratio ($\frac{R_i}{R_m^o}$) and the aspect ratio ($\frac{H}{d}$). The axial position of the helical path ($z_0$) at a fixed azimuthal location ($\theta$) is obtained by the following equation,

$$z_0(\theta) = \frac{P*\lambda}{2\pi}\theta. \tag{1}$$

The radius of the inner cylinder is constant along the axial and azimuthal directions. The radius of the outer cylinder with the helical corrugated surface varies along the axial direction and azimuthal direction. At each $z_0$ position along the helical path, the equation of the radius of the outer cylinder ($r_o(z,\theta)$) and mean radius ($R_m^o$) are given by the following equations,

$$r_o(z,\theta) = R_m^o + asin\left[\frac{2\pi}{\lambda}(z_0(\theta) + z)\right] \tag{2}$$

$$R_m^o = \frac{R_{max} + R_{min}}{2} \tag{3}$$

where $r_o(z,\theta)$ is the radius of the outer cylinder at any axial position, $R_m^o$ is the mean radius of the outer cylinder, $R_{max}$ is the maximum radius of the outer cylinder, $R_{min}$ is the minimum radius of the outer cylinder and $\theta$ is the azimuthal angle.

The inner cylinder is rotating with an angular speed of $\Omega$ and the outer cylinder is fixed. The Reynolds number ($Re$) is defined by the equation below,

$$Re = \frac{\Omega R_i d}{\upsilon}. \tag{4}$$

The pitch or inclination angle of the helical-corrugated surface is expressed using the following equation,

$$\gamma = tan^{-1}\left(\frac{H}{2\pi R_m^o}\right). \tag{5}$$

*Figures 2b-d* illustrate TCF$_\text{Helical}$ for the three values of $P^*$. From the projected view (or plan view) indicated in *Figure 2*, the inner and outer cylinders are concentric in TCF with longitudinal corrugated surfaces (TCF$_\text{Longitudinal}$) (*see Figure 2a with $P^*$=0*) and the helical corrugated surface (*see Figures 2b-d with $P^*$ =1,2,3, respectively*). To satisfy the assumption of the infinite length of the cylinder, periodic boundary conditions are imposed at both ends of the cylinders. The cross section in the r-θ plane between point "M" and "N" indicated in *Figures 2b-d* is used as the computational height ($H$=8$d$) of the cylinder. From the geometric variation of the cross-sections in the r-θ plane of TCF with the longitudinal corrugated surface, the gap between two cylinders is constant at a fixed axial position (*see Figure 2a*). On the other hand, the cross-section (i.e., in r-θ plane) for the helical corrugated surface is not uniform but varies with the pitch to wavelength ratio (see *Figures 2b-d*).

The shape of the cross-section is fixed for each pitch to wavelength ratio but the orientation of the "star-like" cross-section changes with the axial height. Thus, the radial gap between two cylinders varies with both the axial height and azimuthal angle for the helical corrugated surface (see *Figures 2b-d*).

## 3. Numerical configuration

The fluid flow is considered as viscous and incompressible. The inner wall is subjected to a constant angular speed and a no-slip boundary condition is applied at the outer wall. The periodic boundary condition is imposed at the axial ends of both cylinders. Similar to the LES of TCF$_\text{Longitudinal}$ reported in Razzak et al. (2020), the three dimensional Navier-Stokes equations applicable in the TCF$_\text{Helical}$ is solved by LES with the Wall-Adapting Local Eddy-viscosity (WALE) SubGrid Scaling model (SGS) using the pimpleFoam solver in OpenFOAM 5. The numerical schemes and solver configuration used here are exactly the same as that used for LES of TCF with the longitudinal corrugated surface by Razzak et al. (2020). The Courant–Friedrichs–Lewy (CFL) < 0.5 is used for the whole range of $Re$ to ensure stability, and the normalized torque is calculated using the following equation (Fazel & Booz, 1984),

$$Normalized\ Torque, \bar{\tau} = \frac{T}{\nu \rho \Omega R_1^2 H}. \qquad (6)$$

From the transient behaviour of normalized torque for $Re \lesssim 425$ for $A* = 0.25$ and $P* = 1$ at $H = 8d$ (see *Figure 3*), the statistical mean torque becomes fairly independent of time after $t \gtrsim 1$ sec. This indicates that the flow has reached the statistically stationary state after $t \gtrsim 1$ sec, similar to the simulation of TCF$_{Longitudinal}$ by Razzak et al. (2020). Therefore, the simulation time (i.e., flow time) of 5 sec used in the present study is deemed sufficient for the flow structures to reach a statistically stationary state.

Also similar to TCF$_{Longitudinal}$ simulated by Razzak et al. (2020), the structural mesh is generated using Fluent Mesher with the growth rate of the grid size from the wall set at 5 in TCF with the helical corrugated surface. The number of grid points selected along the radial, axial and azimuthal directions are summarized in Table 1 which is similar to that of TCF$_{Longitudinal}$ as indicated in Razzak et al. (2020).

Table 1: Number of grid points along radial, azimuthal and axial directions for each $H = 2d$.

| $Re$ | Radial direction | Azimuthal direction | Axial direction (for each $H = 2d$) |
|---|---|---|---|
| $60 \leq Re \leq 175$ | 70 | 250 | 125 |
| $175 \leq Re \leq 275$ | 85 | 300 | 150 |
| $275 \leq Re \leq 475$ | 100 | 350 | 175 |
| $475 \leq Re \leq 650$ | 115 | 400 | 200 |

Razzak et al. (2020) reported that for TCF$_{Longitudinal}$, the torque and flow structures become independent of height for cylinder with $H \geq 8d$. To verify the dependency of results on the height of cylinders for TCF with helical corrugated surface, several cases of numerical simulation has been performed for $H=2d$, $4d$, $6d$ and $8d$. From the results presented in *Figure 4*, torque is found to be independent of height of cylinder for $H \geq 6d$. Therefore, similar to TCF$_{Longitudinal}$ reported in Razzak et al. (2020), $H=8d$ has been selected as the nominal height of cylinder for TCF with helical corrugated surface in the present study.

Though previously the LES of TCF$_{Longitudinal}$ has been validated for $H = 8d$ against DNS and experiments in Razzak et al. (2020), three cases of DNS at $Re = 175$, 375 and 475 of TCF with the helical corrugated surface for $P* = 1$, $A* = 0.25$ and $H = 8d$ have been performed to further validate the present LES results. Three dimensional Navier-Stokes equations of TCF with the helical corrugated surface has been solved by DNS using the icoFOAM solver in OpenFOAM 5. The mesh configuration and numerical schemes used are the same as the DNS simulation conducted by Razzak et al. (2019) where TCF with smooth walls (TCF$_{Smooth}$) have been studied. The normalized torque ($\bar{\tau}$) obtained in LES exhibits excellent agreement with the DNS results (see *Table 2*) with a maximum of 0.91% difference, giving confidence that the LES used in the present study is able to capture the flow physics sufficiently accurately.

*Table 2: Comparison of $\bar{\tau}$ obtained in LES and DNS for $P* = 1$, $A* = 0.25$.*

| $Re$ | $\bar{\tau}_{LES}$ | $\bar{\tau}_{DNS}$ | Percentage difference |
|---|---|---|---|
| 175 | 25.173 | 25.339 | 0.65% |
| 375 | 35.481 | 35.163 | 0.90% |
| 475 | 38.919 | 38.568 | 0.91% |

# 4. Results and Discussion

The LES results obtained for the three pitch to wavelength ratio ($P*$) of 1, 2, 3 and amplitude to wavelength ratio ($A*$) of 0.25 are described in the following sections.

## 4.1 Flow regimes

To examine the influence of the helical corrugated surface on the oscillating axial secondary flow, a similar analysis to Razzak et al. (2020) is adopted. The normalized area average axial velocity ($\bar{V}_z(t,\theta)$) in the $r$-$Z$ plane at a fixed $\theta$ location and the normalized amplitude of area average axial velocity ($\bar{V}_{z_{rms}}(\theta)$) have been calculated using the following equations,

$$\bar{V}_z(t,\theta) = \frac{\int_{R_i}^{R_o}\int_0^H V_z(r,z,t,\theta)dzdr}{(R_o - R_i)\Omega R_i H} \tag{7}$$

$$\bar{V}_{z_{rms}}(\theta) = \frac{1}{\Omega R_i}\sqrt{\frac{\sum(\bar{V}_z(t,\theta) - mean\bar{V}_z(\theta))^2}{N}} \tag{8}$$

where $V_z(r, z, t, \theta)$ is the axial velocity and $N$ is the number of samples.

From the transient behaviour of the axial velocity $\bar{V}_z(t,\theta)$, for $P* = 1$ and $A* = 0.25$ (see *Figure 5*), the axial velocity $\bar{V}_z(t,\theta)$ is found to be steady at $Re = 100$ when the flow reaches a statistically stationary state (see *Figure 5 a*). At $Re = 125$ and 175, the axial velocity $\bar{V}_z(t,\theta)$ is found to be unsteady (see *Figures 5 b-c*). From the behaviour of the mean axial velocity $\bar{V}_z(\theta)$ with $Re$ at $\theta = 0$ and 180 (mean $\bar{V}_z(\theta)$ has been calculated when the flow structures have reached statistically stationary state), the mean $\bar{V}_z(\theta)$ is found to be non-zero for $P* = 1, 2$ and 3 and $A* = 0.25$ for all $Re$ (see *Figure 6*). The non-zero mean $\bar{V}_z(\theta)$ is broadly classified as the axial secondary flow similar to TCF with longitudinal corrugated surface(i.e., TCF$_{Longitudinal}$) as reported in Razzak et al. (2020). Therefore, based on the transient behaviour of $\bar{V}_z(t,\theta)$ and mean $\bar{V}_z(\theta)$, primarily two types of flow are observed in TCF with helical corrugated surface(i.e., TCF$_{Helical}$). The first flow regime is classified to be steady axial secondary flow and the second one is named as the unsteady axial secondary flow.

In the steady axial secondary flow regime, based on the behaviour of the streamline pattern at $\theta = 0$ and $\theta = 180$ shown in Figure 7a, and the absolute helicity and azimuthal vorticity ($\omega_\theta$) contour at the inner wall, a single stationary vortex along the helical path is observed in the maximum gap region. This flow regime is named as the stationary helical vortex flow (SHVF). The range of $Re$ at which the flow remains as SHVF is found to be $Re \lesssim 100$ for $P*=1$.

The unsteady axial secondary flow is divided into two sub-regimes based on the transient behaviour of $\bar{V}_z(t,\theta)$ presented in *Figures 5 b-c*. In the 1ˢᵗ sub-regime, $\bar{V}_z(t,\theta)$ is found to be periodic at $Re =125$ (see *Figure 5 b*). In this flow regime, the periodic formation of a single vortex at the inner wall and another vortex at the outer wall and their subsequent disappearance result in the formation of periodic oscillating axial secondary flow with an azimuthal wave which can be seen in the azimuthal vorticity contour plotted at the inner wall as shown in *Figure 7 b*. Based on the transient behaviour of $\bar{V}_z(t,\theta)$ and flow structures, the 1ˢᵗ sub-regime of unsteady axial secondary flow is named the periodic helical wavy vortex flow (PHWVF)*(see Figure 7 b)*. The range of $Re$ at which the flow remains as PHWVF is $100 < Re \lesssim 165$ for $P* =1$. In the PHWVF flow regime, a gradual increase in the mean $\bar{V}_z(\theta)$ and its amplitude ($\bar{V}_{z_{rms}}(\theta)$) of oscillation with $Re$ is observed for all the values of $P*$ (*see Figure 6* and *Figure 8*). Furthermore, the flow structures involved in PHWVF are related to one frequency which increases with the increase in $Re$ (*see Figure 9*). The periodic behaviour of oscillating axial secondary flow in PHWVF is very similar to the periodic oscillation of axial secondary flow reported in Razzak et al. (2020) for TCF$_{Longitudinal}$. It can be noted that the periodic oscillation of axial secondary flow occurs at an earlier $Re$ in TCF$_{Helical}$ compared to TCF$_{Longitudinal}$ reported in Razzak et al. (2020).

In the 2ⁿᵈ sub-regime of unsteady secondary axial flow, $\bar{V}_z(t,\theta)$ is found to be non-periodic as $Re$ is increased beyond 165 for $P*=1$ (see *Figure 5 c*). In this flow regime, a pair of vortices at the inner wall and another single vortex at the outer wall are found to occur non-periodically (see streamline pattern associated with flow structures at $\theta = 0$ and $\theta = 180$ shown in *Figure 7c*) and this results in the formation of oscillating axial secondary flow with modulated wave along the azimuthal direction (see azimuthal vorticity contour plotted at inner wall indicated in *Figure 7c*). This flow regime is classified as non-periodic helical wavy vortex flow (NPHWVF). An overall increasing trend in mean $\bar{V}_z(\theta)$ (see *Figure 6*) and a slight decreasing trend in the amplitude ($\bar{V}_{z_{rms}}(\theta)$) (see *Figure 8*) of the non-periodic oscillating axial secondary flow with change in *Re* are found in the NPHWVF flow regime. For *Re* < 215, the dominant flow structure is related to vortices of two frequencies whereas for *Re* >215, the flow structure related to three frequencies (*see Figure 9*). It is also found that the magnitude of all frequencies increases with *Re*. This flow regime is found to be similar to non-periodic oscillating axial secondary flow with azimuthal wave (NANPSAF) reported in TCF$_{Longitudinal}$ by Razzak et al. (2020). As similar to the PHWVF, NPHWVF flow also occurs at an earlier *Re* in TCF$_{Helical}$ compared to TCF$_{Longitudinal}$.

From the above discussion, three flow regimes are clearly observed and delineated for *P*\* =1 and *A*\* = 0.25, with *Re* ranging between 60 to 650. For *P*\* = 2 and 3 at *A*\* = 0.25, stationary helical vortex flow (SHVF) is found to occur at *Re* ≲ 90 for *P*\* =2 and *Re* ≲ 85 for *P*\*=3. The flow structure is found to be periodic helical wavy vortex flow (PHWVF) for *P*\* = 2 with *Re* ranging between 90 to 145, and for *P*\* = 3 with *Re* ranging between 85 to 140. The range of *Re* at which the flow behaves as non-periodic helical wavy vortex flow (NPHWVF) are *Re* > 145 and 140 for *P*\* =2 and 3, respectively. The summary of the range of *Re* for the above described three flow regimes and their dependency on *P*\* at *A*\* = 0.25 is presented in *Table 3*.

*Table 3: The summary of dependency of flow regimes on Re and P\**

| | Steady axial secondary flow | Unsteady Oscillating axial secondary flow | |
|---|---|---|---|
| | | Periodic oscillating axial secondary flow | Non-periodic oscillating axial secondary flow |
| *P*\* | Stationary helical vortex flow (SHVF) | Periodic helical wavy vortex flow (PHWVF) | Non-periodic helical wavy vortex flow (NPHWVF) |
| 1 | *Re* ≲ 100 | 100 < *Re* ≲ 165 | *Re* > 165 |
| 2 | *Re* ≲ 90 | 90 < *Re* ≲ 145 | *Re* > 145 |
| 3 | *Re* ≲ 85 | 85 < Re ≲ 140 | *Re* > 140 |

In the following sections, the typical behaviour of stationary helical vortex flow (SHVF), periodic helical wavy vortex flow (PHWVF) and non-periodic helical wavy vortex flow (NPHWVF) regimes and their corresponding influence on drag are described.

## 4.2 Influence of flow regimes on drag reduction

This section explains the influence of the above described three flow regimes on drag reduction. To examine the influence of the above-mentioned flow regimes on drag reduction, normalized torque ($\bar{\tau}$) (see eqn. 6) and normalized azimuthal vorticity ($\bar{\omega}_\theta$) obtained in TCF with the helical corrugated surface (TCF$_{Helical}$) for the three values of *P*\* (1, 2 and 3) is compared against $\bar{\tau}$ and $\bar{\omega}_\theta$ obtained in TCF with smooth surface (TCF$_{Smooth}$)reported in Razzak et al. (2019) for the same radius ratio (i.e., radius ratio 0.5) and axial height (i.e., *H = 8d*) (*see Figure 10 a*). The drag reduction and area average normalized azimuthal vorticity ($\bar{\omega}_\theta$) are obtained by the following equations,

$$\Delta\,\bar{\tau} = \frac{\bar{\tau}_{TCF\,with\,helical\,corrugated\,outer\,wall} - \bar{\tau}_{TCF\,with\,smooth\,outer\,wall}}{\bar{\tau}_{TCF\,with\,smooth\,outer\,wall}} \times 100\% \qquad (9)$$

$$\bar{\omega}_\theta(t,\theta) = \frac{\int_{R_i}^{R_o}\int_0^H |\omega_\theta|(r,z,t,\theta)dzdr}{(R_o - R_i)\Omega H} \qquad (10)$$

where $\omega_\theta$ is azimuthal vorticity.

The results of drag reduction ($\Delta\bar{\tau}$) vs $Re$ for all three $P^*$ values can be broadly divided into three regions based on the influence of three flow regimes (*see Figure 10 c*). In Region I (SHVF), no drag reduction ($\Delta\bar{\tau} \geq 0$) is found for $Re \leq 85$ (for brevity, see *Figure 10c provided for P\*=1),* beyond this $Re$, drag reduction is observed. When $Re < 85$, $\bar{\omega}_\theta$ is higher for flow with $\text{TCF}_{\text{Helical}}$ compared to $\text{TCF}_{\text{Smooth}}$ (see *Figure 11* shown for $P^*=1$) but this observation reverses for $Re$ >85 to 100 and $\bar{\omega}_\theta$ for $\text{TCF}_{\text{Helical}}$ becomes lower than that for $\text{TCF}_{\text{Smooth}}$ . In the PHWVF flow regime (region II, $100 < Re < 165$) with maximum drag reduction occurring at the upper bound of this $Re$ range ($Re \approx 165$). At the same time within this regime, $\bar{\omega}_\theta$ is lower for $\text{TCF}_{\text{Helical}}$ than for $\text{TCF}_{\text{Smooth}}$ (see *Figure 11* shown for $P^*=1$). The drag reduction ($\Delta\bar{\tau}<0$) in this region is attributed to the emergence of oscillating periodic axial secondary flow observed in the PHWVF flow regime (see *Figure 7 b* and *Figure 10 c*). In region III (NPHWVF), $\Delta\bar{\tau}$ varies continuously depending on the non-periodic oscillating axial secondary flow observed in this flow regime. In Region III, drag reduction is observed for $Re \lesssim 450$ for $P^* =1$, 2 and $Re \lesssim 440$ for $P^* = 3$ as well as for $Re \gtrsim 555$, 560 and 570 and $\bar{\omega}_\theta$ is found to be lower for $\text{TCF}_{\text{Helical}}$ than for $\text{TCF}_{\text{Smooth}}$ (see *Figure 10*). For $450 \lesssim Re \lesssim 550$ however, no drag reduction is observed and $\bar{\omega}_\theta$ is higher for $\text{TCF}_{\text{Helical}}$ than for $\text{TCF}_{\text{Smooth}}$ (see *Figure 11*). For each flow regime, drag reduction is observed in the presence of axial secondary flow, suggesting that the suppression of azimuthal vortices by the oscillating axial secondary flow. The suppression mechanism of azimuthal vorticities induced by oscillating axial secondary flow and their corresponding influence on drag reduction regions in each flow regime are described in the following section in greater detail.

### 4.2.1    Steady axial secondary flow in SHVF

As discussed above, the emergence of steady axial secondary flow in the stationary helical vortex flow (SHVF) regime results in higher drag for $Re <85$, beyond which drag reduction is observed. In this section, the evolution of SHVF and its corresponding influence on drag reduction is discussed.

#### 4.2.1 (a) Transient evolution of SHVF

At $Re = 90$ at $P^* = 1$ and $A^* = 0.25$ as shown in *Figure 12*, SHVF is found to be the result of a flow separation region occurring at the outer wall of the minimum gap region due to the presence of the helical corrugated surface (see *Figure 12 a*). This leads to the formation of a single secondary vortex at the outer wall of the minimum gap region (see *Figure 12 b*) which then gains momentum from the fluid adjacent to the inner wall. Over time, this vortex develops into a helically shaped vortex due to the presence of the helical corrugated surface (see *Figures 12 c-d*) and shifts towards the maximum gap region. This flow is related to the steady axial secondary flow where a single helical vortex at the maximum gap region remains stationary and does not change with time. Therefore, this flow is classified as a stationary helical vortex flow (SHVF). A similar mechanism is present in the SHVF regime at values of $P^*$.

The steady behaviour of the SHVF flow regime observed in $\text{TCF}_{\text{Helical}}$ is similar to the steady flow regimes (ASSWIV and ASSV) reported in Razzak et al. (2020) for $\text{TCF}_{\text{Longitudinal}}$ and axisymmetric Taylor vortex flow (ATVF) reported in Razzak et al. (2019) for $\text{TCF}_{\text{Smooth}}$ . However, some distinct differences are observed between these TCF configurations involving the emergence of an axial secondary flow and flow structures. Most importantly, for $\text{TCF}_{\text{Smooth}}$ and the $\text{TCF}_{\text{Longitudinal}}$, the magnitude of axial secondary flow is zero but it is non-zero for $\text{TCF}_{\text{Helical}}$ (see *Figure 6* and *Figure 13*). Also, the number of vortices for $\text{TCF}_{\text{Helical}}$ in the annular space between two cylinders drops to half that of $\text{TCF}_{\text{Smooth}}$ or a $\text{TCF}_{\text{Longitudinal}}$ for the same cylinder axial height (*see Figure 13*).

#### 4.2.1 (b) Influence of Steady axial secondary flow (SHVF) on drag reduction

As indicated above, no drag reduction ($\Delta\bar{\tau} > 0$) is observed in SHVF flow regime until $Re \lesssim 85$ for the three values of $P^*$ *(see Figures 10 a-b). Figure 10a* indicates that $\bar{\tau}$ for $\text{TCF}_{\text{Helical}}$ remains constant until about $Re = 100$ for the three $P^*$ considered in this study. In this range of $Re$, the magnitude of steady axial secondary flow also remains unchanged with $Re$ (see *Figure 6*) which results in both $\bar{\omega}_\theta$ and $\bar{\tau}$ to be nearly constant (see *Figure 11* shown for $P^*=1$) within the present of range of $Re$ considered for $\text{TCF}_{\text{Helical}}$.

As reported in Razzak et al. (2019) for TCF$_{Smooth}$, laminar Couette flow (i.e., steady flow with no axial secondary flow and no Taylor vortex) is observed until $Re = 68$. This is accompanied by a nearly constant $\bar{\tau}$ until $Re = 68$ (*see Figure 10 a*). In this range of $Re$ ($Re \lesssim 68$), though steady axial secondary flow is observed in TCF$_{Helical}$, $\bar{\omega}_{\theta}$ is found to be higher in TCF$_{Helical}$ compared to TCF$_{Smooth}$ (see *Figure 11*) due to the emergence of stationary helical vortex at the maximum gap region (see *Figure 12 c*) causing increased drag in TCF$_{Helical}$ (*see Figure 10 c*). As $Re$ is increased beyond 68 in TCF$_{Smooth}$, a gradual increase in $\bar{\omega}_{\theta}$ (see *Figure 11*) is observed due to the development of Taylor vortices (see *Figure 13* provided for $Re = 90$ for steady flow with no axial secondary flow) which results in the continuous increase in the shear stress $\bar{\tau}$ (see *Figure 10 a*). However, as described above, despite the development of helical vortices shown in *Figure 13 c* in this range of $Re$, $\bar{\omega}_{\theta}$ remains nearly constant with $Re$ together with a nearly constant steady axial secondary flow magnitude (see *Figure 6*). This in turn results in an almost constant shear stress $\bar{\tau}$ for TCF$_{Helical}$ (see *Figure 10 a*). Therefore, the continuous increase in $\bar{\tau}$ with $Re$ due to the increase in strength of Taylor vortex (rapid increase in $\bar{\omega}_{\theta}$ shown in *Figure 11*) in TCF$_{Smooth}$ results in the decrease in the positive value of $\Delta\bar{\tau}$ (i.e., the difference of $\bar{\tau}$ between TCF$_{Helical}$ and TCF$_{Smooth}$) until $Re <$ 85 (see *Figure 10 b*). Eventually as $Re$ increases above 85, $\bar{\omega}_{\theta}$ for TCF$_{Helical}$ becomes lower than that for TCF$_{Smooth}$ (*see Figure 11*) as the axial secondary flow develops. The number of vortices that develops in TCF$_{Helical}$ is observed to be half the number of vortices observed in TCF$_{Smooth}$ in this flow regime (see *Figures 13 a and c*). This results in a lower wall shear stress in TCF$_{Helical}$ compared to that for TCF$_{Smooth}$ (*see Figure 13 b*), ie. a drag reduction (see *Figure 10 b*). Based on above discussion, it can be concluded that the axial secondary flow that develops in the SHVF regime suppresses the strength of the helical vortices and contributes to drag reduction.

### 4.2.2 Periodic oscillating axial secondary flow in PHWVF

As indicated in *Section 4.1*, increasing $Re$ beyond about 90 for the range of $P^*$ currently considered leads to the development of a single periodic vortex at the inner wall and another vortex at the outer wall. Their subsequent disappearance results in the emergence of a periodic oscillating axial secondary flow and an azimuthal wave. This flow is classified as a periodic helical wavy vortex flow (PHWVF). In this flow regime, periodic oscillating axial secondary flow results in an increase in the drag reduction. In this section, the periodic behaviour of the oscillating axial secondary flow, its corresponding influence on flow structures and drag reduction are discussed in detail.

### 4.2.2 (a) Typical transient evolution of PHWVF

To better understand the behaviour of the oscillating axial secondary flow and its corresponding influence on the flow structure, the typical transient evolution of the PHWVF at $Re = 125$ is presented in *Figure 14* showing the instantaneous streamlines in the r-Z plane of $\theta = 180$ for a single period after the flow has reached a statistically stationary state. As shown in *Figures 14 a-b*, the emergence of a single vortex (first vortex) possibly due to centrifugal instability at the inner wall and its growth results in the increase in the magnitude of the axial secondary flow ($\bar{V}_z(t,\theta)$) (*see Figure 14 k*). This results in the suppression of a pair of vortices of the maximum gap region (*see Figure 14 a-b*) which in turn leads to the decrease in the magnitude of the azimuthal vorticity ($\bar{\omega}_{\theta}(t,\theta)$) (*see Figure 14 l*) and azimuthal wall shear stress (*see Figure 14 m*). As the pair of vortices at the maximum gap region completely disappears (*see Figure 14 c*), the magnitude of the axial secondary flow $\bar{V}_z(t,\theta)$ reaches its maximum value and the magnitude of the azimuthal vorticity $\bar{\omega}_{\theta}(t,\theta)$ reaches its minimum (*see Figure 14 k*). Correspondingly, the wall shear stress reaches its minimum value (*see Figure 14 m*). Over time, as another newly formed single vortex grows further and reaches to its optimal size (*see Figures 14 d-f*), $\bar{V}_z(t,\theta)$ approaches to its minimum value (*see Figure 14 k*) and $\bar{\omega}_{\theta}(t,\theta)$ its maximum value (*see Figure 14 l*). The wall shear stress (*see Figure 14 m*) correspondingly approaches peak value.

As the flow develops, a second single vortex appears and grows in the minimum gap region of the outer wall (see *Figure 14 g*). This second vortex then moves along the axial direction under the influence of the axial secondary flow and stays at the maximum gap region with the previously formed first vortex as a pair of vortices (see *Figure 14 h*). During this stage of flow, $\bar{V}_z(t,\theta)$ increases and $\bar{\omega}_{\theta}(t,\theta)$ decreases, causing the wall shear stress to decrease. As time progresses, another single vortex which is similar to the first vortex appears and grows along the inner wall (see *Figures 14 i-j*). This results in the suppression and disappearance of the pair of vortices at the maximum gap region similar to that described in *Figures 14 a-d* and this is again attributed to the $\bar{V}_z(t, \theta)$ and $\bar{\omega}_\theta(t, \theta)$ attaining their maximum and minimum values respectively. This emergence of a single vortex at the inner wall, followed by another vortex at the outer wall and their suppression and subsequent disappearance occurs periodically. This periodic behaviour of these flow structures results in the periodic oscillating axial secondary flow with azimuthal wave. This suppresses the azimuthal vortices resulting in the reduction of the wall shear stress.

A similar mechanism for the formation of periodic helical wavy vortex flow (PHWVF) is also observed for all other values of $P*$ considered in the present study. The $Re$ range of PHWVF is found to be $100 < Re \lesssim 165$ for $P*=1$, $90 < Re \lesssim 145$ for $P*=2$ and $85 < Re \lesssim 140$ for $P*=3$ which suggests that increasing $P*$ results in the formation of PHWVF flow at an earlier $Re$. At the same time, the maximum $Re$ which flow exhibits PHWVF also decreases with an increase in $P*$.

The periodic behaviour of oscillating axial secondary flow (PHWVF) observed in TCF$_{Helical}$ is found to be similar to the axisymmetric oscillating periodic axial secondary flow (APSAF) reported by Razzak et al. (2020) for TCF$_{Longitudinal}$ where the periodic formation of a pair of vortices at the minimum gap region and their subsequent disappearance results in the formation of oscillating axial secondary flow (APSAF). This type of flow is very similar to that of TCF$_{Helical}$ where the periodic formation and disappearance of the single vortex at the inner wall and another vortex at the outer wall contributes to the occurrence of oscillating axial secondary flow. Though the oscillating axial secondary flow is periodic in both TCF configurations, the flow structures are axisymmetric with no azimuthal wave in TCF$_{Longitudinal}$ which is related to the non-axisymmetric periodic flow with the azimuthal wave in TCF$_{Helical}$. Besides, the periodic oscillating axial secondary flow occurs at an earlier $Re$ in TCF$_{Helical}$ than that of TCF$_{Longitudinal}$.

### 4.2.2 (b) Influence of periodic oscillating axial secondary flow (PHWVF) on drag reduction

As discussed in *section 4.1*, drag in the PHWVF flow regime of TCF$_{Helical}$ is found to be lower ($\Delta\bar{\tau} < 0$) than TCF$_{Smooth}$ (*see Figures 10 a-b*) for $P* =1$ and $Re$ ranging between 100 to 165. A similar behaviour is observed for $P* =2$ (i.e., $90 < Re \lesssim 145$) and $P* =3$ (i.e., $85 < Re \lesssim 140$). In these ranges of $Re$ for periodic helical wavy vortex flow (PHWVF) (see typical *Figure 7 b*), an increase in the oscillating axial secondary flow in terms of both the mean and amplitude with $Re$ is observed (see *Figure 6, Figure 8* and *see Figure 7 b*). This is true for all the three values of $P*$. On the other hand, the flow structures in TCF$_{Smooth}$ as reported in Razzak et al. (2019) was found to be steady and axisymmetric with no axial secondary flow (i.e., mean $\bar{V}_z = 0$) (*see Figure 13 a*). As shown in *Figure 11*, in the same range of $Re$, $\bar{\omega}_\theta$ is found to be lower in TCF$_{Helical}$ than that of TCF$_{Smooth}$. An emergence of oscillating axial secondary flow in TCF$_{Helical}$ results in the suppression of strength of azimuthal vortices in this flow regime. It was also shown in the previous section (i.e., *section 4.2.2(a)*) that the suppression of azimuthal vortices under the influence of oscillating axial secondary flow results in the decrease in the wall shear stress (*see Figures 14 k-m*). It can also be seen that suppression of azimuthal vorticities under the influence of oscillating axial secondary flow results in lower wall shear stress in TCF$_{Helical}$ compared to that of TCF$_{Smooth}$ (*see Figures 14 k-m*). This emergence of oscillating axial secondary flow and its corresponding suppression of strength of vortices (*see Figure 11 and Figure 14*) thus contributes to drag reduction. This is in agreement with the findings of Razzak et al. (2019) where the emergence of oscillating axial secondary flow in TCF$_{Smooth}$ for $Re > 425$ was found to suppress the strength of vortices (i.e., azimuthal vorticity) and resulted to a sudden decrease in torque *(see Figure 10 a)*. Moreover, the behaviour of oscillating axial secondary flow in PHWVF and its contribution in drag reduction may bear possible similarity to the drag reduction via spanwise wall oscillation (eg. Mangiavacchi et al., 1992; Laadhari et al., 1994; Baron & Quadrio, 1996; Choi & Graham, 1998; and Choi & Clayton, 2001; Skote, 2011;2012; Hehner et al., 2019; Yao et al., 2019). In these studies, spanwise wall oscillation was found to suppress the strength of streamwise vortices (i.e., streamwise vorticity) as well as the pushing away of these vortices away from wall, leading to a thickening of the viscous layer and causing the wall shear stress to decrease. In line with the above-mentioned findings, the formation of oscillating axial secondary flow (i.e., non-zero mean $\bar{V}_z$ as shown in *Figure 6*) may also contribute to the thickening of the viscous layer by suppressing the strength of vortices (i.e., azimuthal vorticity), thereby resulting in lower drag in TCF$_{Helical}$ compared to TCF$_{Smooth}$ at the corresponding $Re$ range with no secondary axial flow. In this flow regime, as $Re$ increases, the difference of $\bar{\tau}$ between TCF$_{Helical}$ and TCF$_{Smooth}$ ($\Delta\bar{\tau}$) continues to increase for all the three values of $P*$ (see *Figure 10 b*). As discussed above, the mean and amplitude of oscillating axial secondary flow increases with $Re$ and this results in further suppression of the vorticities (i.e., azimuthal vorticity) and thereby contributes to the increase in the magnitude of $\Delta\bar{\tau}$ in the PHWVF flow regime.

### 4.2.3 Non-Periodic oscillating axial secondary flow in NPHWVF

As shown in *section 4.2*, in the NPHWVF flow regime, non-periodic oscillating axial secondary flow results in the occurrence of drag reduction only when the magnitude of the azimuthal vorticity is smaller in TCF$_{Helical}$ compared to that of TCF$_{Smooth}$ (*see Figure 10 b* and *Figure 11*). Therefore, to better understand the influence of non-periodic oscillating axial secondary flow on drag reduction, this section examines the typical transient evolution of non-periodic oscillating axial secondary flow in NPHWVF flow regime and its corresponding influence on the suppression of azimuthal vortices.

### 4.2.3 (a) Transient evolution NPHWVF

From the typical change in the instantaneous behaviour of streamline patterns in the *r-Z* plane of $\theta = 0$ and $\theta = 180$ for $Re$ =175, initially, a single vortex appears at the inner wall of the minimum gap region (*see Figures 15 a-b*) following which another single vortex forms at the inner wall near the maximum gap region (*see Figure 15 c*). The formation of the first vortex may be related to the flow separation occurring at the minimum gap region and the second one could be related to centrifugal instability. This new pair of vortices increases in size with time and weakens an old pair of vortices in the maximum gap region *(see Figures 15 d-e)*. The second vortex strengthens further and takes over the place of the previous pair of vortices of the maximum gap region *(see Figures 15 e-g)*. At the same time, the first vortex becomes weaker and disappears from the minimum gap region of the outer wall (*see Figures 15 f-h*). Following this, another pair of vortices appears at the inner wall ( *see Figures 15 h-i*) and this also results in the weakening and disappearance of the vortices in the maximum gap region ( *see Figures 15 h-i*). Over time, in addition to the pair of vortices at the inner wall, another single vortex is found to form at the minimum gap region at the outer wall (*see Figure 15 i*). The formation of a pair of vortices at the inner wall, another single vortex at the outer wall and their subsequent influence on the weakening and disappearance of vortices of the maximum gap region occur non-periodically. This flow phenomenon results in the formation of a non-periodic oscillating axial secondary flow with modulated azimuthal wave which can be seen in the azimuthal vorticity contour plotted at the inner wall presented in *Figure 16 a*. In this flow regime, a pair of vortices at the inner wall and a single vortex at the outer wall were found to occur non-periodically until $Re$ =215. However, as the $Re$ is increased beyond 215, the non-periodic formation of three vortices was found to occur in the inner wall along with vortices of the outer wall. This results in the modulation of the azimuthal wave being more prominent and can be seen in the azimuthal vorticity contours plotted at the inner wall for $Re$ =225 and 325, indicated in *Figures 16 b-c*. A similar observation in the occurrence of non-periodic helical wavy vortex flow (NPHWVF) is also found for all the values of $P*$. The flow regime is found to be NPHWVF at $Re$ >165, 145 and 140 for $P*$ =1, 2, 3, respectively, and suggests that increasing $P*$ results in the formation of NPHWVF flow at an earlier $Re$.

The influence of the non-periodic formation of pairs of vortices and their subsequent disappearance in the oscillating axial secondary flow with the modulated wave in the NPHWVF flow regime for TCF$_{Helical}$ is similar to the formation of non-periodic and non-axisymmetric oscillating axial secondary flow with azimuthal wave (NANPSAF) reported by Razzak et al. (2020) for TCF$_{Longitudinal}$. In that reported flow regime, the non-periodic formation of a pair of vortices at the minimum gap region resulted in the formation of oscillating axial secondary flow with azimuthal wave. However, non-periodic oscillating axial secondary flow in TCF$_{Helical}$ is found to occur at an earlier $Re$ than that of TCF$_{Longitudinal}$. Moreover, the mean flow velocity of oscillating axial secondary flow in TCF$_{Helical}$ is far greater than that of TCF$_{Longitudinal}$.

### 4.2.3   (b) Influence of non-periodic oscillating axial secondary flow (NPHWVF) on drag reduction

Based on the comparison of $\Delta\bar{\tau}$ between TCF$_{Helical}$ and TCF$_{Smooth}$, the region III (see *Figure 10 c shown for P*=1*) in the NPHWVF flow regime where $Re \gtrsim 165$, 145 and 140 for the three values of *P**, respectively, can be broadly further divided into three sub-regions (see *Figure 10 c*). In sub-region "A", the flow in TCF$_{Smooth}$ was reported to be axisymmetric and steady with no-secondary flow (i.e., mean $\bar{V}_z = 0$) (*see Figure 13 a and Razzak et al., 2019*) but the TCF$_{Helical}$ exhibits unsteady oscillating axial secondary flow with non-zero mean $\bar{V}_z$ (*see Figure 10 c and Figure 5 b for Re =175*) and results in a lower $\bar{\omega}_\theta$ for TCF$_{Helical}$ compared to that of TCF$_{Smooth}$ (*see Figure 11*). This suggests that the oscillating axial secondary flow with non-zero mean $\bar{V}_z$ suppresses the strength of azimuthal vortices and contributes to drag reduction (*see Figure 10 b*) in this sub-region "A" (i.e., $165 \lesssim Re \lesssim 450$ for *P** =1, $145 \lesssim Re \lesssim 450$ for *P** =2 and $140 \lesssim Re \lesssim 440$ for *P** =3). This is in agreement with the finding of PHWVF flow regime, where the oscillating axial secondary flow with non-zero mean $\bar{V}_z$ suppresses the strength of vortices, leading to drag reduction. In the sub-region "A", an increase in $Re$ results in a mildly perceptible decrease in the value of $\Delta\bar{\tau}$ until $Re = 215$ for the three values of *P** (see *Figure 10 b*). In this range of $Re$, despite a gradual increase in the magnitude of axial secondary flow until $Re = 215$ (see *Figure 6*), the increasing rate of the mean and amplitude of oscillating axial secondary flow decreases with $Re$  (*see Figure 6 and Figure 8*) due to the inability of the additional pair of vortices at the inner wall (*see Figure 15 and Figures 16 b- c*) to significantly suppress the strength of the vortices. This may be related to the decrease in drag difference between TCF$_{Helical}$ and TCF$_{Smooth}$ (i.e., a decrease in the magnitude of  $\Delta\bar{\tau}$) with $Re$ for the three values of *P**.  From $Re$ ranging between 215 to 450, 450 and 440 for *P**=1,2,3, respectively, the mean and amplitude of oscillating axial secondary flow either increases slightly or decreases (see *Figure 6 and Figure 8*) under the influence of the formation of three or more vortices at the inner wall which suggests that the oscillating secondary flow becomes too weak to suppress the strength of vortices with the increase in $Re$. This results in an overall decreasing trend in $\Delta\bar{\tau}$ in the sub-region "A" .

In sub-region " B" (i.e., $450 < Re \lesssim 555$ for *P** =1,  $450 < Re \lesssim 560$ for *P** =2 and $440 < Re \lesssim 570$ for *P** =3), drag in TCF$_{Helical}$ is found to be higher ($\Delta\bar{\tau} > 0$) than TCF$_{Smooth}$ (see *Figure 10 c*). In this sub-region, oscillating axial secondary flow is observed in both TCF$_{Helical}$ and TCF$_{Smooth}$ (see Razzak et al., 2019). In TCF$_{Smooth}$, a sudden increase in the magnitude of the oscillating axial secondary flow with $Re$ was found to suppress the strength of Taylor vortices leading to the sudden drop in drag with $Re$ (see *Figure 10 a and Figure 11*) (*Razzak et al., 2019*). Though the magnitude of the oscillating axial secondary flow is found to be larger in TCF$_{Helical}$ compared to TCF$_{Smooth}$, the increasing trend with $Re$ in the magnitude of the oscillating axial secondary flow for TCF$_{Helical}$ is observed to be very small (see *Figure 6*) for the three values of *P** and may result in a weaker suppression of the strength of vortices. This causes the magnitude of $\bar{\omega}_\theta$ to be larger in TCF$_{Helical}$ compared to that of TCF$_{Smooth}$ (*see Figure 11*). This is considered the primary cause of higher drag ($\Delta\bar{\tau} > 0$) occurring in TCF$_{Helical}$ compared to that of TCF$_{Smooth}$ in the sub-region "B".

In the sub-region "C" (i.e., $Re > 555$, 560 and 570 for the three values of *P**, respectively), the drag in TCF$_{Helical}$ is found to be lower than TCF$_{Smooth}$ (see *Figure 10 b*). In this sub-region, the mean flow velocity of the oscillating axial secondary flow is found to be higher in TCF$_{Helical}$ (see *Figure 6*). Besides, in the same range of $Re$, the mean flow velocity of the oscillating axial secondary flow was found to decrease with $Re$ in TCF$_{Smooth}$ (*Razzak et al., 2019*). These results in a lower magnitude of $\bar{\omega}_\theta$ in TCF$_{Helical}$ compared to that of TCF$_{Smooth}$, and contributing to drag reduction in the sub-region "C".

### 4.3   The comparison of drag reduction between TCF$_{Helical}$ and TCF$_{Longitudinal}$

This section explains the comparison of drag reduction ($\Delta\bar{\tau}$) reported in Razzak et al. (2020) for TCF$_{Longitudinal}$ (*P** =0 and A* = 0.25, 0.2149 and 0.1875) and TCF$_{Helical}$ (i.e., *P**=1, 2 and 3 at A* = 0.25) (*see Figure 10*). To better understand influence of these two types of corrugated surface on drag reduction, a comparison of drag reduction between TCF$_{Helical}$ and TCF$_{Longitudinal}$ are presented in *Figure 10 c* for *P\* =0 and A\* = 0.25 (*TCF$_{Longitudinal}$*) and *P\* =1 and A\* = 0.25* (TCF$_{Helical}$).

As shown in Figure 10 c, for *85 < Re <100*, in the SHVF flow regime, drag reduction is observed in TCF$_{Helical}$ (see *Figure 10c*). However, no drag reduction is found in TCF$_{Longitudinal}$ for *Re ≲ 105* (see *Figure 10 c*). For this range of *Re*, a non-zero steady axial secondary flow is observed in TCF$_{Helical}$ while a steady axisymmetric flow with no secondary flow is observed for TCF$_{Longitudinal}$ (see *Figures 13 c-d*). This results in a lower magnitude of $\overline{\omega}_\theta$ in TCF$_{Helical}$ (*see Figure 11*) compared to TCF$_{Longitudinal}$. As shown in *Figures 13 c-d*, the formation of axial secondary flow in TCF$_{Helical}$ also results in the number of vortices reducing and dropping to half that of TCF$_{Longitudinal}$. Therefore, the emergence of steady axial secondary flow in TCF$_{Helical}$ suppresses the strength of azimuthal vorticities and leads to drag reduction for *Re* ranging between 85 to 100 (see *Figure 10 c and Figure 13 b*).

For *100 ≲ Re ≲ 165*, an increase in the mean and amplitude of the oscillating axial secondary flow (*see Figure 6 and Figure 8*) with *Re* continues to suppress the strength of azimuthal vortices (see *Figure 11*) and this results in drag reduction in TCF$_{Helical}$ (*see Figure 10 c and Figure 14m*). In TCF$_{Longitudinal}$, the occurrence of a pair of secondary vortices with no secondary flow in the minimum gap region was found to suppress the strength of vortices azimuthal (*see Figure 11*) which leads to drag reduction for 105 ≲ Re ≲ 138 (*see Figure 10 c*). Within this range of *Re*, the mean and amplitude of the oscillating axial flow increases as *Re* increases (*see Figure 6 and Figure 8*), resulting in the a stronger suppression of the azimuthal vortices in TCF$_{Helical}$ compared to TCF$_{Longitudinal}$ (*see Figure 11 and Figure 14 m*). The result is nearly 2 times higher drag reduction in TCF$_{Helical}$ compared with that for TCF$_{Longitudinal}$.

For 165 < Re ≲ 455 for TCF$_{Helical}$ and 138 ≲ Re ≲ 440 for TCF$_{Longitudinal}$, oscillating axial secondary flow is observed in both TCF configurations. Though oscillating axial secondary flow is observed in the mentioned ranges of *Re* in both TCF configurations, the magnitude of oscillating axial secondary flow (non-zero mean $\overline{V}_z$ ) is found to be far smaller in TCF$_{Longitudinal}$ than that in TCF$_{Helical}$. This results in a stronger suppression of azimuthal vortices in TCF$_{Helical}$ (*see Figure 11*) and leads to higher drag reduction in TCF$_{Helical}$ compared to TCF$_{Longitudinal}$ (*see Figure 10 c*).

For 455 < Re ≲ 550 for TCF$_{Helical}$ and 440 ≲ Re ≲ 550 for TCF$_{Longitudinal}$, no drag reduction is observed in these two TCF configurations (*see Figure 10c*). Even though oscillating axial secondary flow is observed in TCF$_{Helical}$, TCF$_{Longitudinal}$ and TCF$_{Smooth}$, the suppression of azimuthal vortices is found to be stronger in TCF$_{Smooth}$ compared to the other two TCF configurations (*see Figure 11*). This contributes to a decrease in drag as *Re* increases for TCF$_{Smooth}$. However, no drag reduction is observed for TCF$_{Helical}$ as well as TCF$_{Longitudinal}$ for the above-mentioned ranges of *Re*.

For Re > 555, the magnitude of oscillating axial secondary flow is again able to suppress the azimuthal vortices (*see Figure 11*) and this again results in drag reduction. In this range of *Re*, drag reduction in TCF$_{Helical}$ is slightly higher than TCF$_{Longitudinal}$ and this may be attributed to a higher magnitude of oscillating axial secondary flow in TCF$_{Helical}$ compared to TCF$_{Longitudinal}$.

In general, the oscillating axial secondary flow is found to occur at an earlier *Re* with a larger mean of axial secondary flow in TCF$_{Helical}$ than that of TCF$_{Longitudinal}$. This results in a stronger suppression of azimuthal vortices in TCF$_{Helical}$ compared to that in TCF$_{Longitudinal}$, leading to drag reduction at an earlier *Re* for TCF$_{Helical}$. In addition, drag reduction observed in TCF$_{Helical}$ is found to be higher compared to TCF$_{Longitudinal}$.

## Conclusion

The drag reduction capability of naturally occurring oscillating axial secondary flow induced by helical corrugated surface in Taylor Couette flow (TCF) set up has been investigated for the three values of the pitch to wavelength ratios ($P^*$) (1, 2 and 3) and amplitude to wavelength ratio ($A^*$) of 0.25. As reported by Razzak et al. (2019;2020) for TCF with smooth and longitudinal corrugated surface, an increase in the magnitude of naturally occurring oscillating axial secondary flow with $Re$ resulted in the emergence of drag reduction. This provides an idea that drag reduction might be enhanced if it is possible to generate stronger oscillating axial secondary flow by the means of manipulation of surface geometry. Therefore, the motivation of present study is to enhance naturally occurring oscillating axial secondary flow induced by helical corrugated surface in TCF and its corresponding influence on drag reduction. From the variation of axial secondary flow with $Re$ for all the three values of $P^*$, it is found that axial secondary flow is observed for $Re$ ranging between 60 to 650. Based on the transient behaviour of axial secondary flow (i.e., when the flow reaches the statistically stationary state), primarily two types of flow are observed in TCF with the helical corrugated surface; they are steady axial secondary flow and unsteady oscillating axial secondary flow. In steady axial secondary flow, a single vortex along the helical path is observed in the maximum gap region at $Re$ = 60 for the three values of $P^*$. This flow regime is classified as stationary helical vortex flow (SHVF). In this flow regime, occurrence of steady axial secondary flow was found to suppress the strength of azimuthal vortices which in turn contributes to the emergence of drag reduction. The unsteady axial secondary flow has been divided into two sub-regimes. In the first sub-regime, the periodic formation of a single vortex at the inner wall and another vortex at the minimum gap region of the outer wall and their subsequent disappearance result in the formation of periodic oscillating axial secondary flow with azimuthal waves for $Re$ = 100, 90 and 85 for the three values of $P^*$, respectively. This flow regime is classified as periodic helical wavy vortex flow (PHWVF). The ranges of $Re$ at which flow remains as PHWVF are $100 \lesssim Re \lesssim 165$ for $P^*$ = 1, $90 \lesssim Re \lesssim 145$ for $P^*$ =2 and, $85 \lesssim Re \lesssim 140$ for $P^*$ = 3. In this flow regime, the occurrence of oscillating axial secondary flow continues to suppress the strength of azimuthal vortices which results in a maximum of 13% drag reduction. In the second sub-regime of unsteady axial secondary flow, the non-periodic formation of a pair of vortices at the inner wall and a single vortex at the outer wall results in the formation of non-periodic oscillating axial secondary flow with modulated waves as $Re$ increases beyond 165, 145 and 140 for the three values of $P^*$, respectively. This flow regime is named as the non-periodic helical wavy vortex flow (NPHWVF). At the presence of non-periodic oscillating axial secondary flow in NPHWVF flow regime, drag reduction is observed only when magnitude of azimuthal vorticities is smaller in TCF with helical corrugated surface than that of TCF with smooth surface. From the comparison of oscillating axial secondary flow between TCF with longitudinal corrugated surface and helical corrugated surface, oscillating axial secondary flow is found to occur at an earlier $Re$ in TCF with helical corrugated surface. In addition, the magnitude of oscillating axial secondary flow was found to be higher in TCF with helical corrugated surface than TCF with longitudinal corrugated surface which resulted in suppression of azimuthal vortices to be stronger in TCF with helical corrugated surface. This resulted in drag reduction observed in TCF with helical corrugated surface to be nearly two times higher than TCF with longitudinal corrugated surface. The oscillating axial secondary flow reported in this study is similar to the application of spanwise wall oscillation studies for drag reduction in general streamwise direction. It can be noted that oscillating axial secondary flow observed in present study occurs naturally which does not require any external energy or a complex system. However, spanwise wall oscillation studies reported for drag reduction requires external energy and a complex system. Therefore, emergence of naturally occurring oscillating axial secondary flow induced by manipulating surface geometry in the form of helical corrugated surface in the present study is a simple and sustainable method and may contribute significantly to future drag reduction studies.

## Acknowledgements

The authors acknowledge the NUS Research Scholarship. The numerical computation has been performed on the resources of High-Performance Computation (HPC) of the National University of Singapore and National Supercomputing Centre, Singapore.

## DATA AVAILABILITY

Raw data were generated at the [High-Performance Computation (HPC) of the National University of Singapore and National Supercomputing Centre, Singapore] large scale facility. Derived data supporting the findings of this study are available from the corresponding author upon reasonable request.

# References

Bauer, G., Gamnitzer, P., Gravemeier, V. & Wall, W. A.,"An isogeometric variational multiscale method for large-eddy simulation of coupled multi-ion transport in turbulent flow". *Journal of Computational Physics, 251*, 194–208 (2013).

Bazilevs, Y. & Akkerman, I., "Large eddy simulation of turbulent Taylor–Couette flow using isogeometric analysis and the residual-based variational multiscale method". *Journal of Computational Physics, 229*, 3402–3414 (2010).

Baron, A., Quadrio, M., "Turbulent drag reduction by spanwise wall oscillations". *Applied Scientific Research*, *55*(311) (1996).

Bhambri, P. & Fleck, B., Drag reduction using high molecular weight polymers in Taylor-Couette flow". *International Journal of Mechanical and Production Engineering Research and Development., 6*(1), 59-72(2016).

Chen, Y., Floryan, J. M., Chew, Y. T., & Khoo, B. C., "Groove-induced changes of discharge in channel flows". *Journal of Fluid Mechanics*, *799*, 297–333(2016).

Choi, K. S., "Near-wall structure of a turbulent boundary layer with riblets. *Journal of Fluid Mechanic".*, *208*, 417-458(1989)..

Choi, H., Moin, P. & Kim, J. "Direct numerical simulation of turbulent flow over riblets". *Journal of Fluid Mechanics, 255*, 503-539(1993).

Choi, K. S., & Graham, M., "Drag reduction of turbulent pipe flows by circular-wall oscillation". *Physics of Fluids*, *10*(1), 7–9(1998).

Choi, K., & Clayton, B. R., "The mechanism of turbulent drag reduction with wall oscillation". *International Journal of Heat and Fluid Flow*, *22*, 1–9 (2001).

Climent, E., Simonnet, M. & Magnaudet, J. "Preferential accumulation of bubbles in Couette-Taylor flow patterns". *Physics of fluids, 19* (2007).

Coustols, E. & Savill, A. M., "*Turbulent Skin-Friction Drag Reduction By Active and Passive Means. Part 1. Everything you wanted to Know about Riblets, LEBUs and Other Devices,"*. France: Office National D'etudes Et De Recherches Aerospatiales Toulouse (France) (1992)..

Donnelly, R. J., "Experiments on the stability of viscous flow between rotating cylinders. I. Torque". *Proceedings of the Royal Society of London. Series A, Mathematical and Physical Sciences, 246*(1246), 312-325 (1958).

DeGroot, C. T., Wang, C., & Floryan, J. M., "Drag Reduction Due to Streamwise Grooves in Turbulent Channel Flow". *Journal of Fluids Engineering, Transactions of the ASME, 138*(12) (2016).

Drozdov, S., Rafique, M., & Skali-Lami, S., "An asymmetrical periodic vortical structures and appearance of the self-induced pressure gradient in the modified Taylor flow". *Theoretical and Computational Fluid Dynamics, 18*, 137-150 (2004).

Dutcher, C. S. & Muller, S. J. , "The effects of drag reducing polymers on flow stability : Insights from the Taylor-Couette problem". *Korea-Australia Rheology Journal, 21*(4), 223-233 (2009).

Eskinn, D., "Applicability of a Taylor–Couette device to characterization of turbulent drag reduction in a pipeline". *Chemical Engineering Science., 116*, 275-283(2014)..

Fasel, H. & Booz, O., " Numerical investigation of supercritical Taylor-vortex flow for a wide gap". *Journal of Fluid Mechanics, 138*, 21-52 (1984).

Ferrante, A. & Elghobashi, S., "On the physical mechanisms of drag reduction in a spatially developing turbulent boundary layer laden with microbubbles". *Journal of Fluid Mechanics, 503,* 345-355(2004).

Fish, F. & Lauder, G., "Passive and active flow control by swimming fishes and mammals". *Annual Review of Fluid Mechanics, 38,* 193-224 (2006).

Floryan, J. M., "Centrifugal instability of Couette flow over a wavy wall". *Physics of Fluids, 14*(1) (2002).

Ghebali, S., Chernyshenko, S. I., & Leschziner, M. A. (2017b). Can large-scale oblique undulations on a solid wall reduce the turbulent drag? *Physics of Fluids, 29*(10). https://doi.org/10.1063/1.5003617

Greidanus, A. J., Delfos, R., Tokgoz, S. & Westerweel, J., "Turbulent Taylor–Couette flow over riblets: drag reduction and the effect of bulk fluid rotation". *Experiments in Fluids, 56*(107) (2015).

García-Mayoral, R. & Jiménez, D. J., "Drag reduction by riblets". *Phil. Trans. R. Soc. A, 369,* 1412–1427(2011).

Groisman, A. & Steinberg, V., "Couette-Taylor Flow in a Dilute Polymer Solution". *Physical Review Letters, 77*(8) (1996).

Gao, X., Kong, B. & Vigil, R. D., "CFD simulation of bubbly turbulent Tayor–Couette flow". *Chinese Journal of Chemical Engineering., 26*(6), 719-727 (2016).

Greidanus, A. J., Delfos, R., Tokgoz, S. & Westerweel, J., "Turbulent Taylor–Couette flow over riblets: drag reduction and the effect of bulk fluid rotation". *Experiments in Fluids, 56*(107) (2015).

Hall, T. & Joseph, D., "Rotating cylinder drag balance with application to riblets". *Experiments in Fluids, 29*(3), 215–227 (2000).

Hehner, M. T., Gatti, D. & Kriegseis, J., "Stokes-layer formation under absence of moving parts - A novel oscillatory plasma actuator design for turbulent drag reduction". *Physics of Fluids, 31*(5) (2019).

Ikeda, E. & Maxworthy, T., "Spatially forced corotating Taylor-Couette flow". *Physical Review E, 49*(6) (1994).

Kalashnikov, V., "Dynamical similarity and dimensionless relations for turbulent drag reduction by polymer additives". *Journal of Non-Newtonian Fluid Mechanics, 75,* 209–230 (1998).

Ketabdari, M. J., & Saghi, H., Large Eddy Simulation of Laminar and Turbulent Flow on Collocated and Staggered Grids. *ISRN Mechanical Engineering*, *2011*, 1–13 (2011).

Koschmieder, E. L., "Effect of finite disturbances on axisymmetric Taylor vortex flow". *Physics of Fluids, 18,* 499–503 (1975).

Koeltzsch, K., Qi, Y., Brodkey, R. & Zakin, J. "Drag reduction using surfactants in a rotating cylinder geometry". *Experiments in Fluids., 34,* 515–530(2003) .

Laadhari, F., Skandaji, L., & Morel, R., "Turbulence reduction in a boundary layer by a local spanwise oscillating surface". *Physics of Fluids*, *6*(10), 3218–3220 (1994).

Lim, T. T. & Tan, K. S., "A note on power-law scaling in a Taylor–Couette flow". *Physics of Fluids, 16*(140) (2004).

Mangiavacchi, N., Jung, W. J. & Akhavan, R., "Suppression of turbulence in wall-bounded flows by high-frequency spanwise oscillations". *Physics of Fluids, 4*(8), 1605–1607 (1992).

Marxen O., Rist U.,"DNS and LES of the Transition Process in a Laminar Separation Bubble". In: Friedrich R., Geurts B.J., Métais O. (eds) *Direct and Large-Eddy Simulation V*. ERCOFTAC Series, vol 9. Springer, Dordrecht (2004).

Mohammadi, A., & Floryan, J. M., "Mechanism of drag generation by surface corrugation". *Physics of Fluids*, *24*(1) (2012).

Mohammadi, A., & Floryan, J. M., "Pressure losses in grooved channels". *Journal of Fluid Mechanics*, *725*, 23–54(2013a).

Mohammadi, A., & Floryan, J. M., "Groove optimization for drag reduction". *Physics of Fluids*, *25*(11) (2013b).

Mohammadi, A., & Floryan, J. M., Numerical analysis of laminar-drag-reducing grooves. *Journal of Fluids Engineering, Transactions of the ASME*, *137*(4) (2015).

Moradi, H. V., & Floryan, J. M., "Flows in annuli with longitudinal grooves". *Journal of Fluid Mechanics*, *716*, 280–315 (2013).

Moradi, H. V., & Floryan, J. M., "Laminar flow in grooved pipes". *AIAA Journal*, *55*(5), 1749–1752 (2017).

Moradi, H. V., & Floryan, J. M., "Drag reduction and instabilities of flows in longitudinally grooved annuli". *Journal of Fluid Mechanics*, *865*, 328–362 (2019).

Murai, Y., Oiwa, H. & Takeda, Y. "Bubble behavior in a vertical Taylor-Couette flow". *Journal of Physics., 14*, 143–156(2005)..

Ng, J. H., Jaiman, R. K. & Lim, T. T., "Interaction dynamics of longitudinal corrugations in Taylor-Couette flows". *Physics of Fluids*, *30*(9) (2018).

Ohsawa, A., Murata, A. & Iwamoto, K., "Through-flow effects on Nusselt number and torque coefficient in Taylor-Couette-Poiseuille flow investigated by large eddy simulation". *Journal of Thermal Science and Technology, 11*(2) (2016).

Painter, B. D. & Behringer, R. P., "Effects of spatial disorder on the transition to Taylor vortex flow", *Europhysics Letters*, *44* (599) (1998).

Paghdar, D., Jogee, S., & Anupindi, K., "Large-eddy simulation of counter-rotating Taylor–Couette flow: The effects of angular velocity and eccentricity". *International Journal of Heat and Fluid Flow*, *81*, 108514 (2020).

Perlin, M., Dowling, D. R. & Ceccio, S. L., "Freeman Scholar Review: Passive and Active Skin-Friction Drag Reduction in Turbulent Boundary Layers". *Journal of Fluids Engineering, 138*(091104-1) (2016).

Poncet, S., Soghe, R. D., Bianchini, C., Viazzo, S. & Aubert, A., "Turbulent Couette–Taylor flows with end wall effects: A numerical benchmark". *International Journal of Heat and Fluid Flow, 44*, 229–238 (2013).

Poncet, S., Viazzo, S. & Oguic, R., "Large eddy simulations of Taylor-Couette-Poiseuille flows in a narrow-gap system". *Physics of Fluids*, 26, 105108 (2014).

Popiolek, T. L., Awruch, A. M., & Teixeira, P. R. F., Finite element analysis of laminar and turbulent flows using LES and subgrid-scale models. *Applied Mathematical Modelling*, *30*(2), 177–199 . (2006).

Quadrio, M. ,"Drag reduction in turbulent boundary layers by in-plane wall motion". *Royal Society*, *369*(1940), 1428–1442(2011).

Quan, V., "Couette flow with particle injection". *International Journal of Heat and Mass Transfer, 15*(11), 2173-2186(1972)..

Razzak, M. A., Khoo, B.C. & Lua, K.B., "Numerical study on wide gap Taylor Couette flow with flow transition". *Physics of Fluids, 31*(11) (2019).

Razzak, M. A., Khoo, B.C. & Lua, K.B., "Numerical study of Taylor Couette flow with longitudinal corrugated surface". *Physics of Fluids, 32(5)* (2020).

Rosenberg, B. J., Buren, T. V. & Fu, M. K., "Turbulent drag reduction over air- and liquid- impregnated surfaces". *Physics of Fluids, 28*(1) (2016).

Salhi, Y., Si-Ahmed, E.-K., Degrez, G. & Legrand, J., "Numerical Investigations of Passive Scalar Transport in Turbulent Taylor-Couette Flows: Large Eddy Simulation Versus Direct Numerical Simulations". *Journal of Fluids Engineering, 134*(4), 041105 (2012).

Singh, N. K., "Large-eddy simulation of a laminar separation bubble". *Journal of Applied Fluid Mechanics*, *12*(3), 777–788 (2019).

Skote, M., "Turbulent boundary layer flow subject to streamwise oscillation of spanwise wall- velocity". *Physics of Fluids, 23*(081703) (2011).

Skote, M., "Temporal and spatial transients in turbulent boundary layer flow over an oscillating wall". *International Journal of Heat and Fluid Flow, 38*, 1–12 (2012).

Srinivasan, S., Kleingartner, J. A., Gilbert, J. B., Cohen, R. E., Milne, A. J. & McKinley, G. H., "Sustainable Drag Reduction in Turbulent Taylor Couette flows". *Physical Review Letters, 114,* 014501 (2014).

Sugiyama, K., Calzavarini, E. & Lohse, D., "Microbubbly drag reduction in Taylor–Couette flow in the wavy vortex regime". *J. Fluid Mech, 608*, 21–41(2008).

Synge, J. L., "On the Stability Of A Viscous Liquid Between Rotating Coaxial Cylinders". *Proceedings of the Royal Society of London. Series A, Mathematical and Physical Sciences, 167*, 250-256 (1938).

Taylor, G. I., "Stability of a Viscous liquid contained between two rotating cylinders". *Philosophical Transactions of the Royal Society of London. Series A, 223*, 289-343 (1923).

Yao, J., Chen, X., & Hussain, F., "Reynolds number effect on drag control via spanwise wall oscillation in turbulent channel flows". *Physics of Fluids, 31*(8) (2019).

Vakarelski, I. U., Marston, J. O., Chan, D. Y. & Thoroddsen, S. T. ,"Drag Reduction by Leidenfrost Vapor Layers". *Phys. Rev. Lett., 106* (2011).

Yadav, N., Gepner, S. W., & Szumbarski, J., "Flow dynamics in longitudinally grooved duct. *Physics of Fluids, 30*(10) (2018).

Yi, M.-K. & Kim, C., "Experimental studies on the Taylor instability of dilute polymer solutions". *Journal of Non-Newtonian Fluid Mechanics, 72*, 113-139 (1997).

Yuan, W., Xu, H., Khalid, M., & Radespiel, R., "A parametric study of les on laminar-turbulent transitional flows past an airfoil". *International Journal of Computational Fluid Dynamics, 20*(1), 45–54 (2006).

Wang, Y., Wang, D., Guo, W., Yin, J. & Hu, Y., "The effect of smaller turbulent motions on heat transfer in the annular". *Annals of Nuclear Energy, 91*, 1–7 (2016).

Walsh, M. J. ,"Riblets as a Viscous Drag Reduction Technique". *AIAA Journal, 21*(4), 485-486(1983).

Watanabe, K. & Akino, T., "Drag Reduction in Laminar Flow Between Two Vertical Coaxial Cylinders". *Journal of Fluids Engineering, 121*(3), 541-547 (1999).

Weickert, M., Teike, G., Schmidt, O., & Sommerfeld, M. "Investigation of the LES WALE turbulence model within the lattice Boltzmann framework". *Computers and Mathematics with Applications*, *59*(7), 2200–2214 (2010).

Zhu, X., Ostilla-Mónico, R., Verzicco, R. & Lohse, D., "Direct numerical simulation of Taylor–Couette flow with grooved walls: torque scaling and flow structure". *Journal of Fluid Mechanics, 794*, 746–774 (2016).

**Figures:**

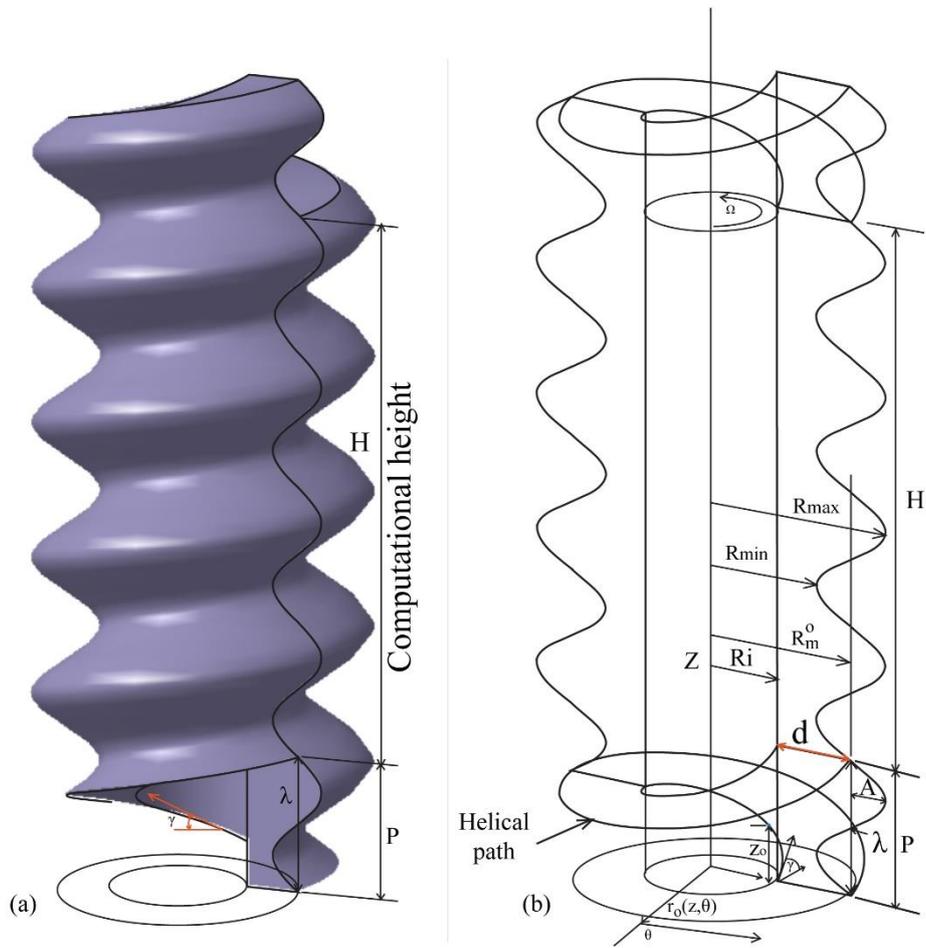

*Figure 1: (a) Shaded view of Taylor Couette flow with a helical-corrugated outer surface (b) Schematic view of Taylor Couette flow with a helical-corrugated surface($TCF_{Helical}$) at the outer cylinder for a pitch to the amplitude ratio of 1(Here, A is the amplitude of corrugated surface, $\lambda$ is wavelength of corrugated surface and P is pitch of helical path).*

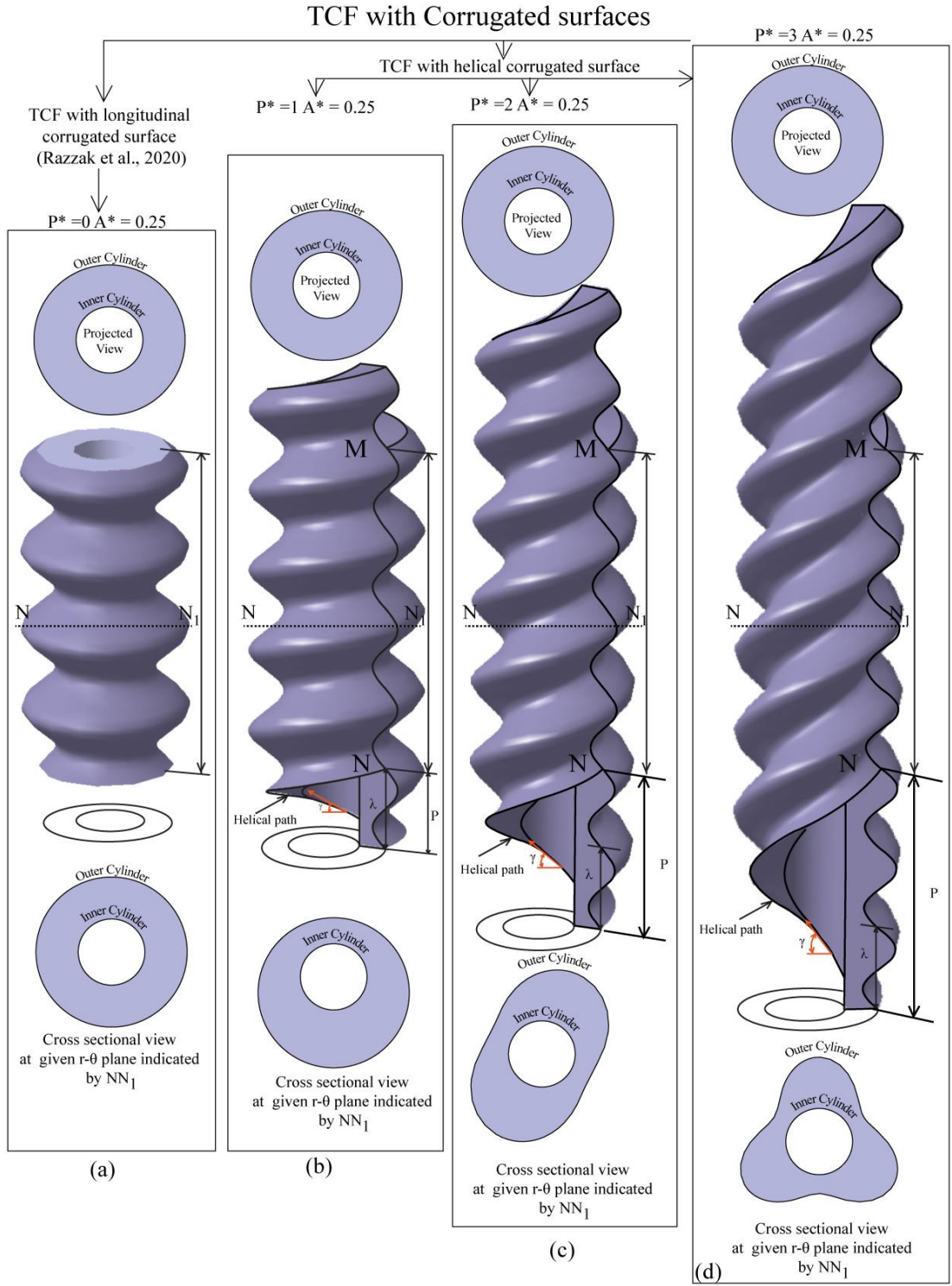

*Figure 2 : The variation of the TCF with helical corrugated surface(TCF$_{Helical}$) with the change in pitch to wavelength ratio (P\*) at A\* = 0.25 (a) TCF with the longitudinal corrugated surface(TCF$_{Longitudinal}$) (reproduced from [Razzak, M. A., Khoo, B.C. & Lua, K.B., "Numerical study of Taylor–Couette flow with longitudinal corrugated surface". Physics of Fluids 32, 053606 (2020)], with the permission of AIP Publishing) (b) TCF with helical corrugated surface (TCF$_{Longitudinal}$) for P\* =1 (c) TCF with helical corrugated surface (TCF$_{Helical}$) for P\* = 2 (d) TCF with helical corrugated surface (TCF$_{Helical}$ ) for P\* = 3.*

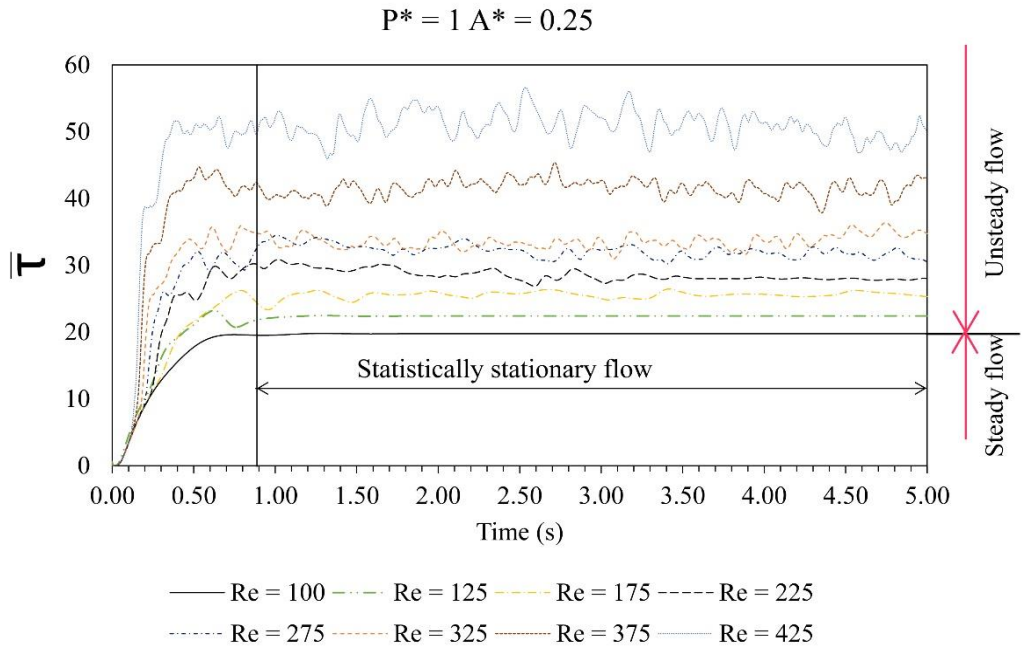

*Figure 3 : Transient behavior of torque for A\* = 0.25 and P\* = 1.*

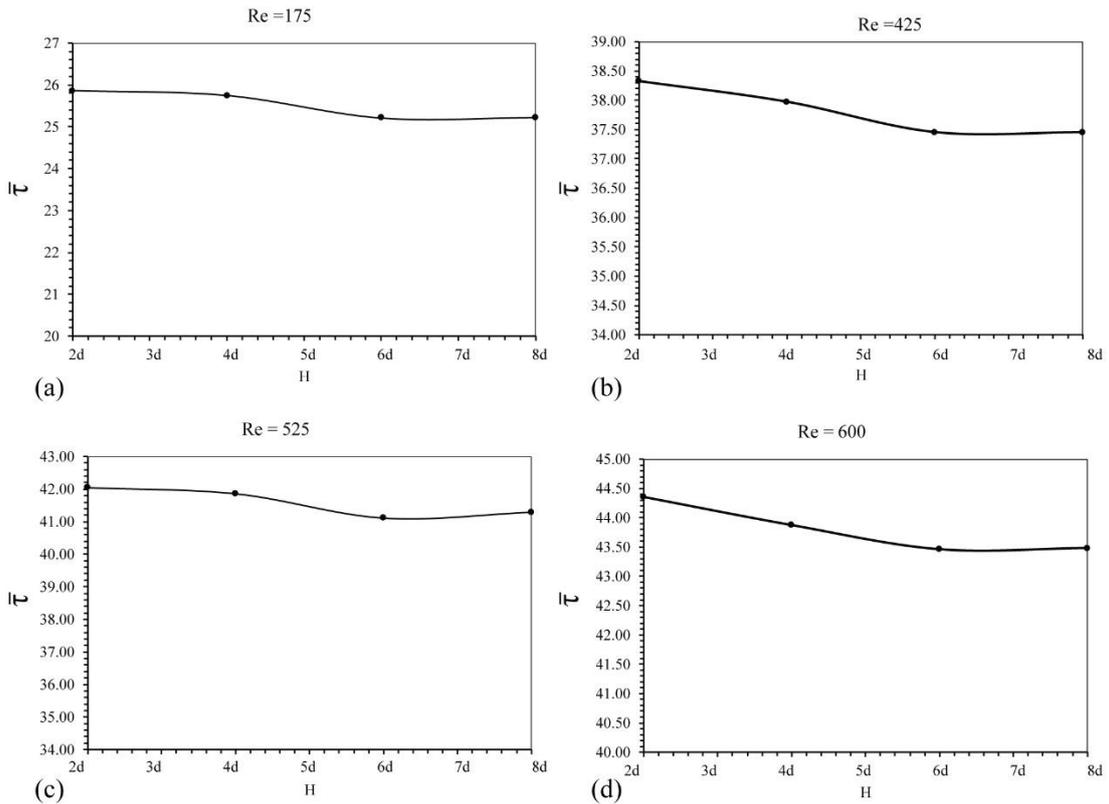

*Figure 4 : Dependency of normalized torque on the height of cylinders (a) dependency of normalized torque on the height of cylinders for Re =175 (b) dependency of normalized torque on the height of cylinders for Re =425(c) dependency of normalized torque on the height of cylinders for Re =525 (d) dependency of normalized torque on the height of cylinders for Re =600.*

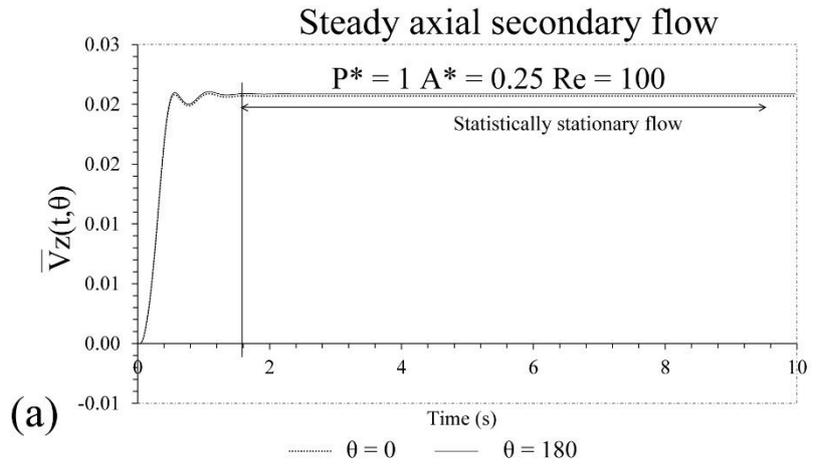

(a)

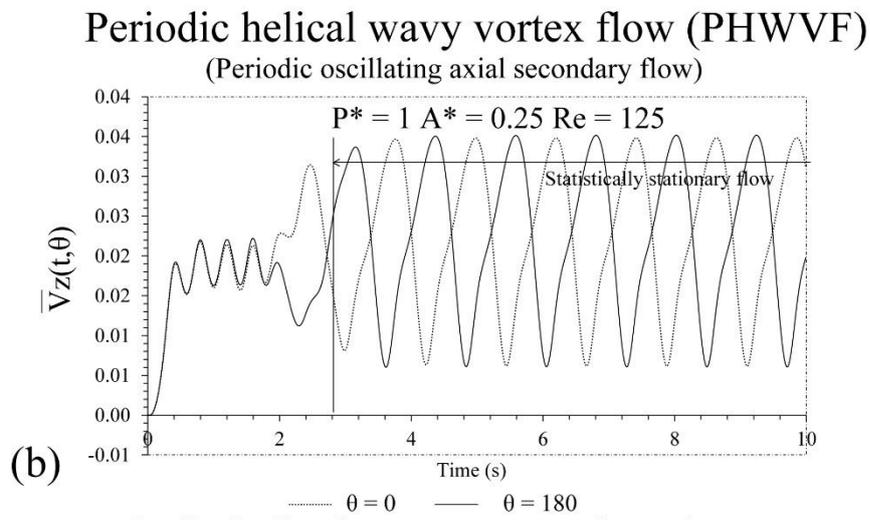

(b)

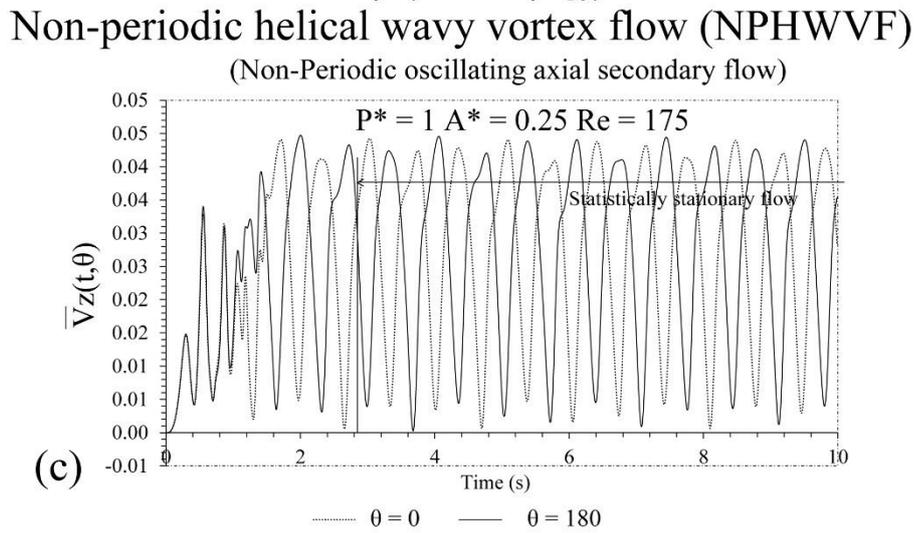

(c)

*Figure 5 : The transient behavior of $\bar{V}_z(t,\theta)$ in the r-Z plane at $\theta$ = 0 and 180 (a) transient behaviour of $\bar{V}_z(t,\theta)$ for Re = 100 (b) transient behavior of $\bar{V}_z(t,\theta)$ at Re = 125(c) transient behavior of $\bar{V}_z(t,\theta)$ at Re = 175.*

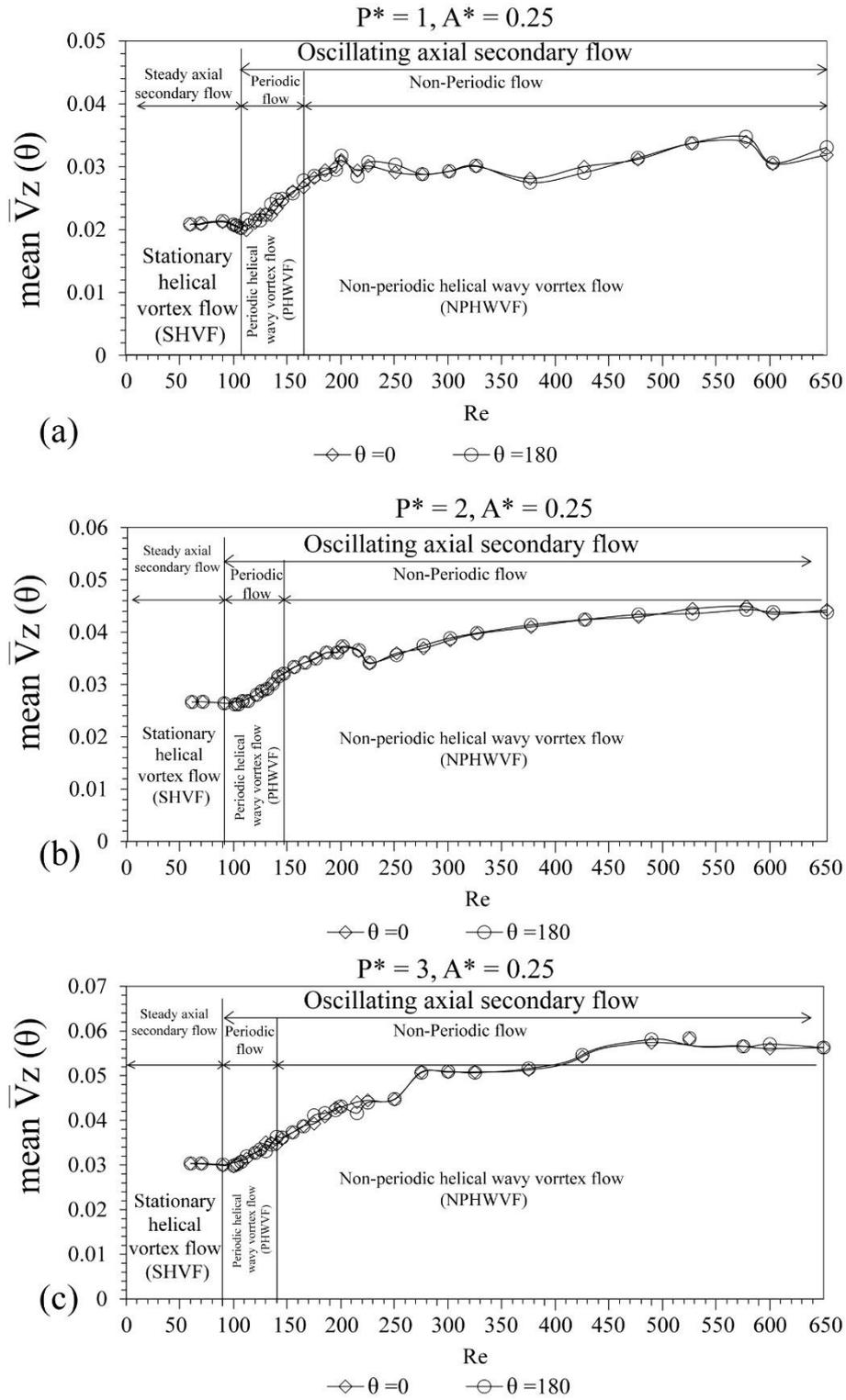

*Figure 6 : The variation of mean $\bar{V}_z(\theta)$ with Re at A\* = 0.25(a) variation of mean $\bar{V}_z(\theta)$ for P\* = 1, (b) variation of mean $\bar{V}_z(\theta)$ for P\* = 2, (c) variation of mean $\bar{V}_z(\theta)$ for P\* = 3.*

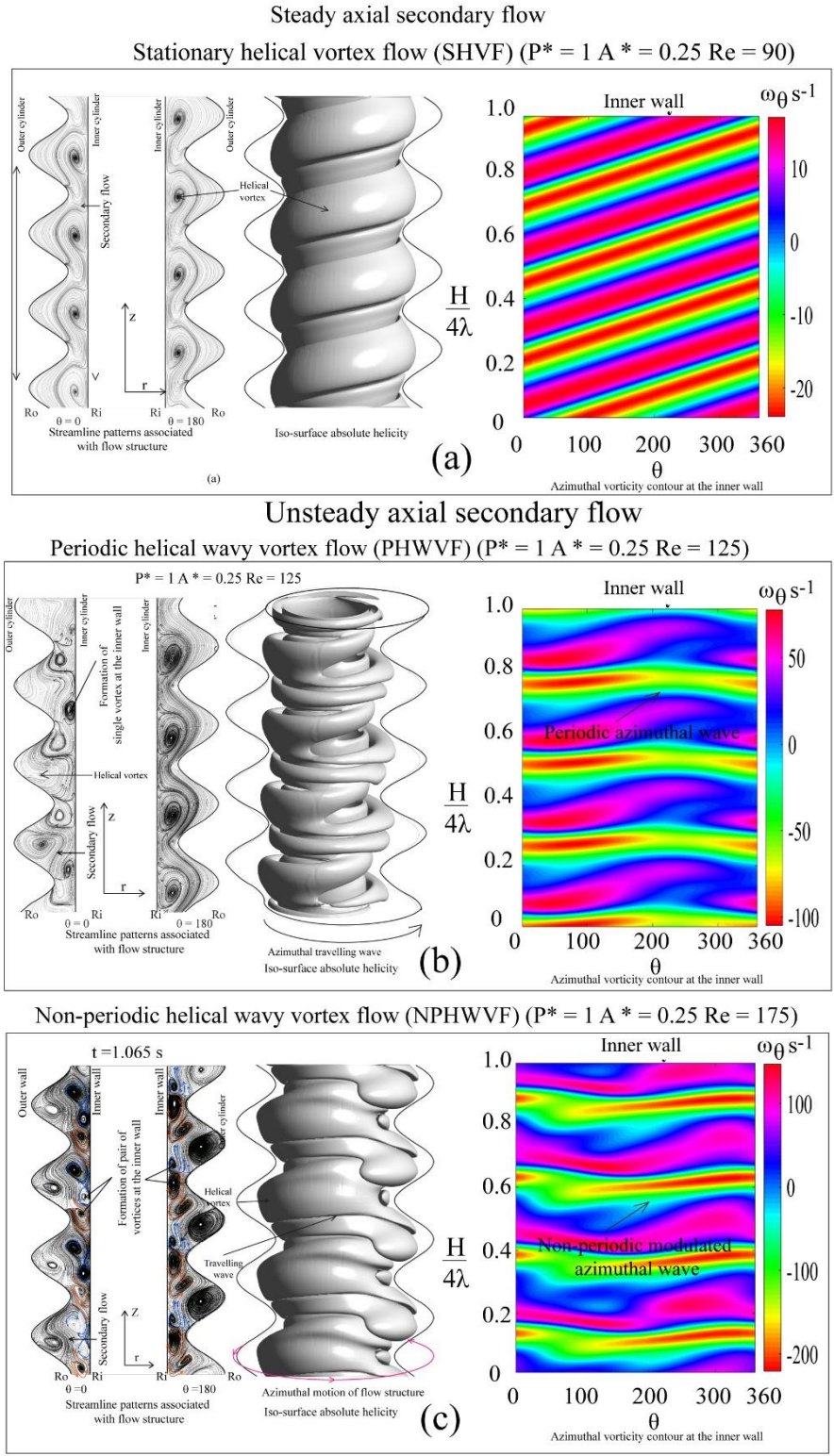

*Figure 7: The streamline patterns, absolute helicity ( i.e., the degree of linkage of the vortex lines of flow which is scalar product of vorticity and velocity vectors and it is conserved when conditions are such that these vortex lines are frozen in the fluid) and azimuthal vorticity contour at the inner wall observed three flow regimes for P\* = 1 and A\* = 0.25(a) stationary helical vortex flow (SHVF) at Re = 90 (time-invariant), (b) typical periodic helical wavy vortex flow (PHWVF) at Re = 125 and t=2.5 s, (c) typical non-periodic helical wavy vortex flow (NPHWVF) at Re = 175 and t=2.5 s.*

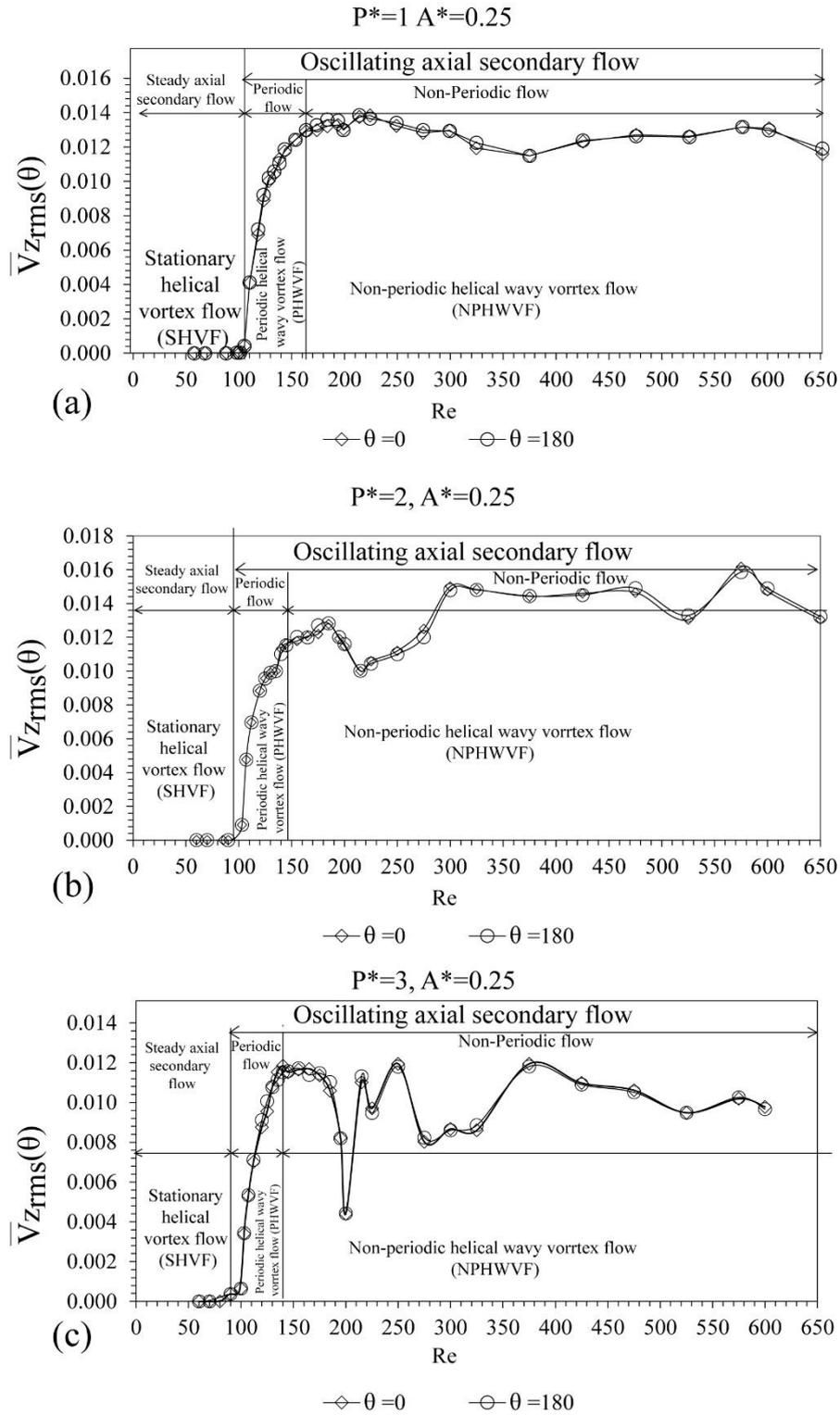

*Figure 8 : The variation of the amplitude of oscillating axial secondary flow ( $\bar{V}_{z_{rms}}(\theta)$ )) with Re at A\* = 0.25(a) variation of $\bar{V}_{z_{rms}}(\theta)$ for P\* = 1, (b) variation of $\bar{V}_{z_{rms}}(\theta)$ for P\* = 2, (c) variation of $\bar{V}_{z_{rms}}(\theta)$ for P\* = 3.*

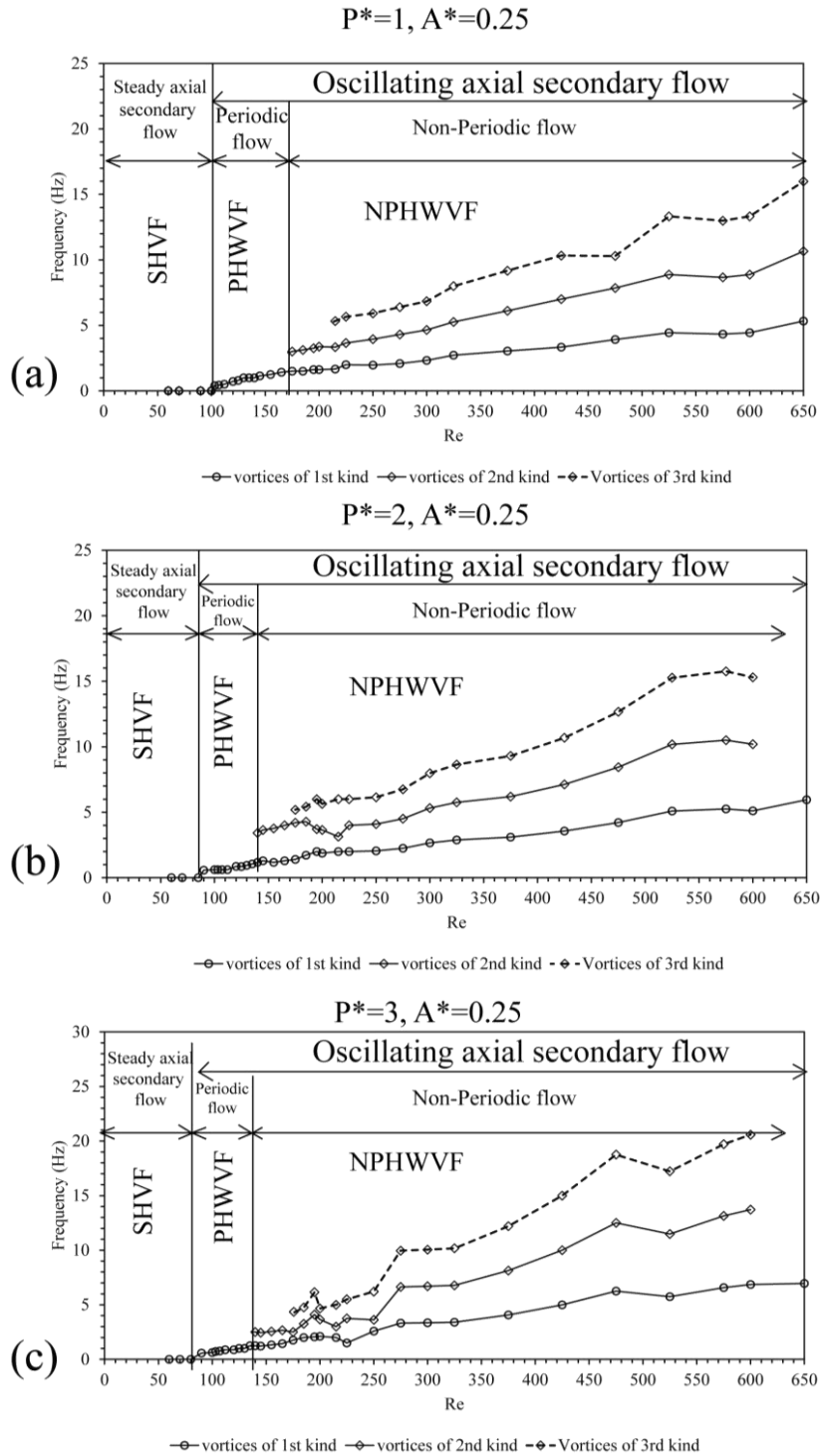

*Figure 9 : The variation of frequency of oscillating axial secondary flow with Re at A\* = 0.25,(a) variation of frequency for P\* = 1, (b) variation of frequency for P\* = 2, (c) variation of frequency for P\* = 3.*

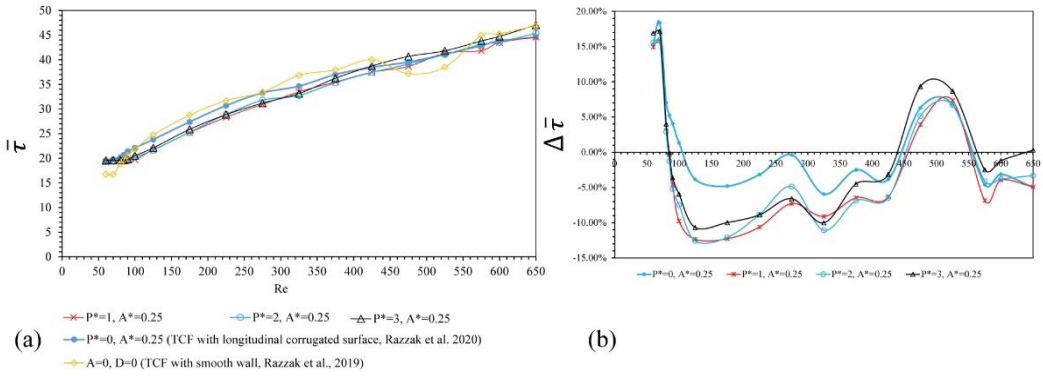

(a)                                                                                      (b)

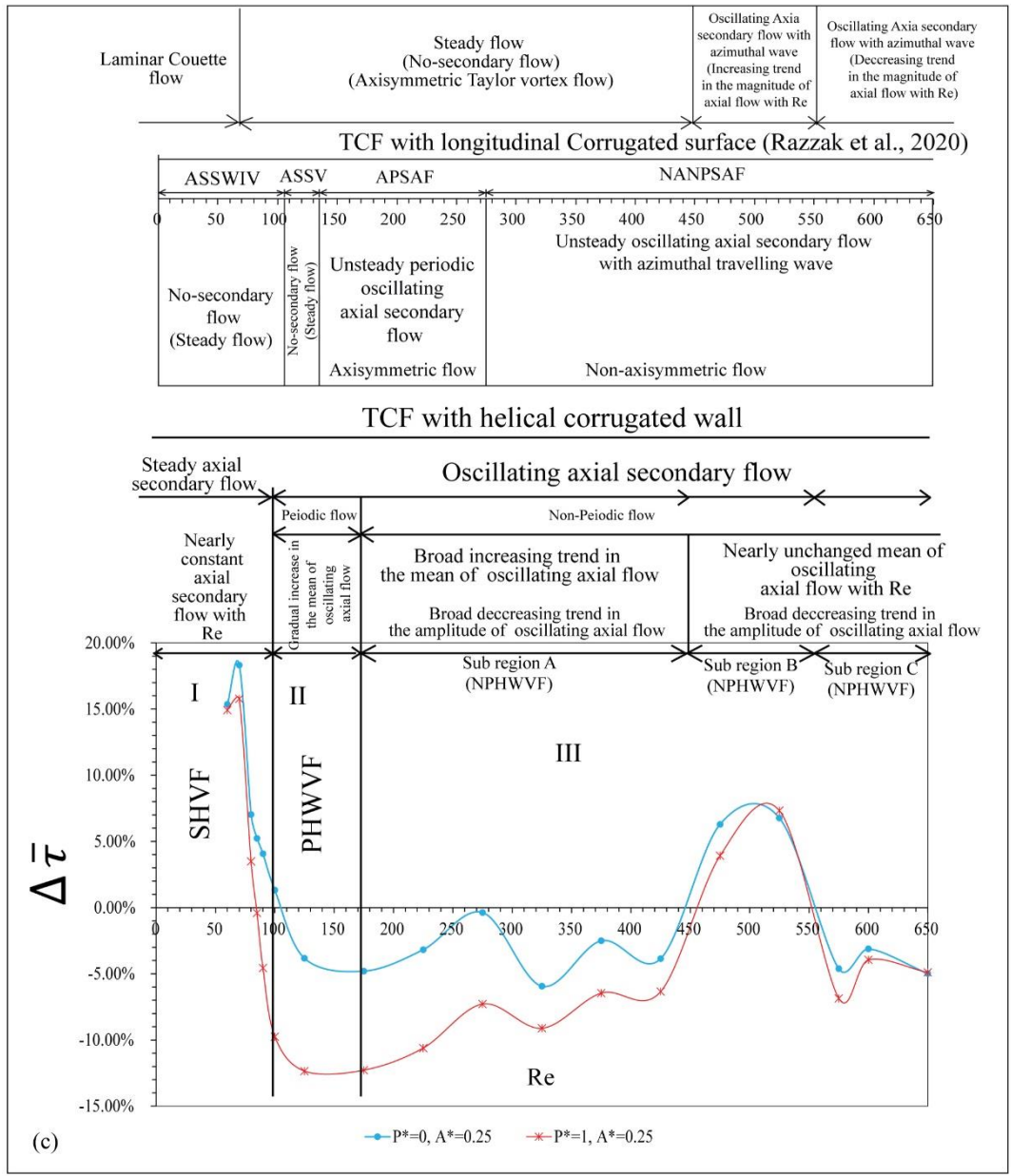

Figure 10: (a) The normalized torque obtained for $A^* = 0, 0.25$ and $P^* = 0, 1,2,3$ (b) the variation of $\Delta\bar{\tau}$ with Re for $A^* = 0.25$ and $P^* = 0,1,2,3$ (c) variation of $\Delta\bar{\tau}$ in SHVF, PHWVF and NPHWVF flow regimes (result presented

*in Figure 10a for TCF with smooth cylinder (A\*=0) is reproduced from [Razzak, M. A., Khoo, B.C. & Lua, K.B., "Numerical study on wide gap Taylor Couette flow with flow transition". Physics of Fluids, 31(11) (2019)] and [Razzak, M. A., Khoo, B.C. & Lua, K.B., "Numerical study of Taylor–Couette flow with longitudinal corrugated surface". Physics of Fluids 32, 053606 (2020)], with the permission of AIP Publishing ).*

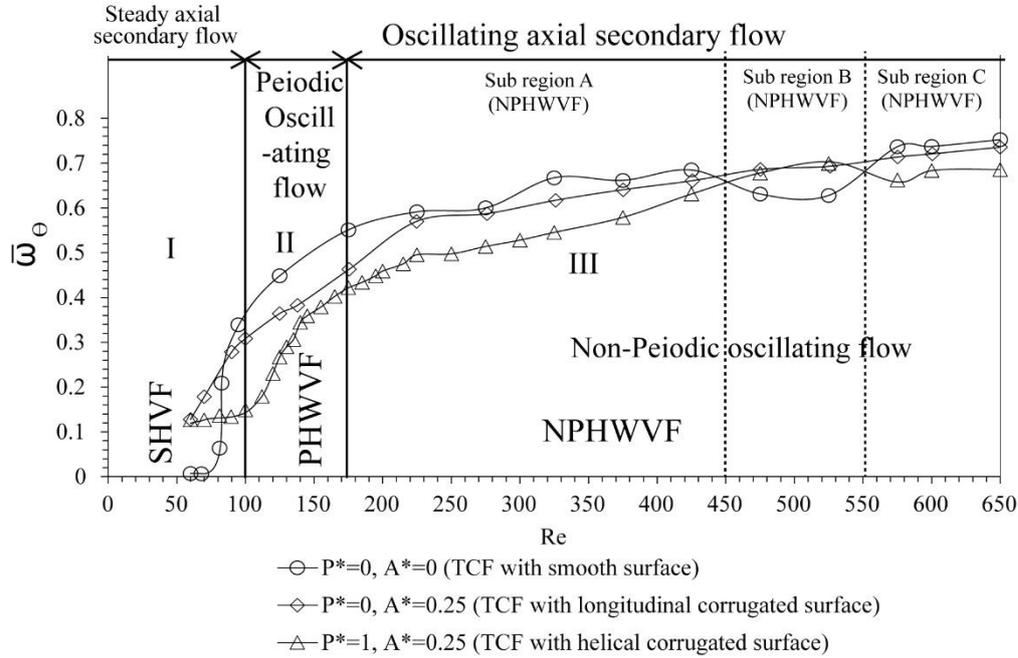

*Figure 11: Variation of normalized azimuthal vorticity ($\bar{\omega}_\theta$) with Re in the r-Z plane of θ=0 (result presented for TCF with smooth cylinder (TCF_{Smooth}) (A\*=0) is reproduced from [Razzak, M. A., Khoo, B.C. & Lua, K.B., "Numerical study on wide gap Taylor Couette flow with flow transition". Physics of Fluids, 31(11) (2019)] and TCF with helical corrugated surface (TCF_{Helical}) is produced from [Razzak, M. A., Khoo, B.C. & Lua, K.B., "Numerical study of Taylor–Couette flow with longitudinal corrugated surface". Physics of Fluids 32, 053606 (2020)], with the permission of AIP Publishing ).*

## Source and propagation of stationary
## helical vortex flow (SHVF)

### P* = 1 A* = 0.25 Re = 90

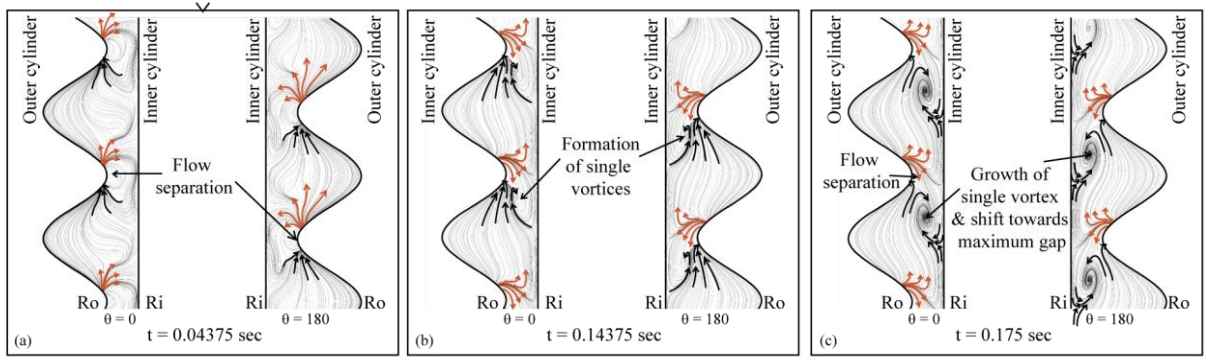

→ Fluid moving away from outer wall
← Fluid moving towards outer wall

### Stationary helical vortex flow (SHVF)

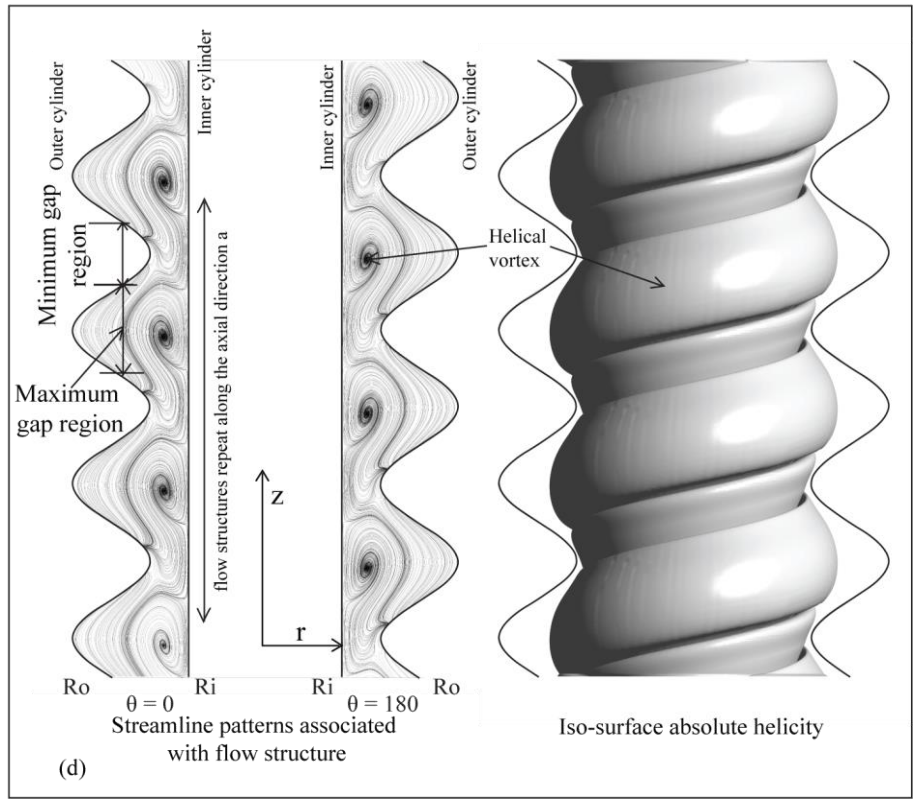

*Figure 12: Streamline patterns associated with the evolution of flow structure of the stationary helical vortex flow (SHVF) for P* =1 and A* = 0.25 at Re = 90 (a) formation of flow separation at the outer wall of minimum gap region (b)formation of single secondary vortex, (c) growth of single secondary vortex, (d) formation of stationary helical vortex flow (SHVF) when the flow reaches the statistically stationary state.*

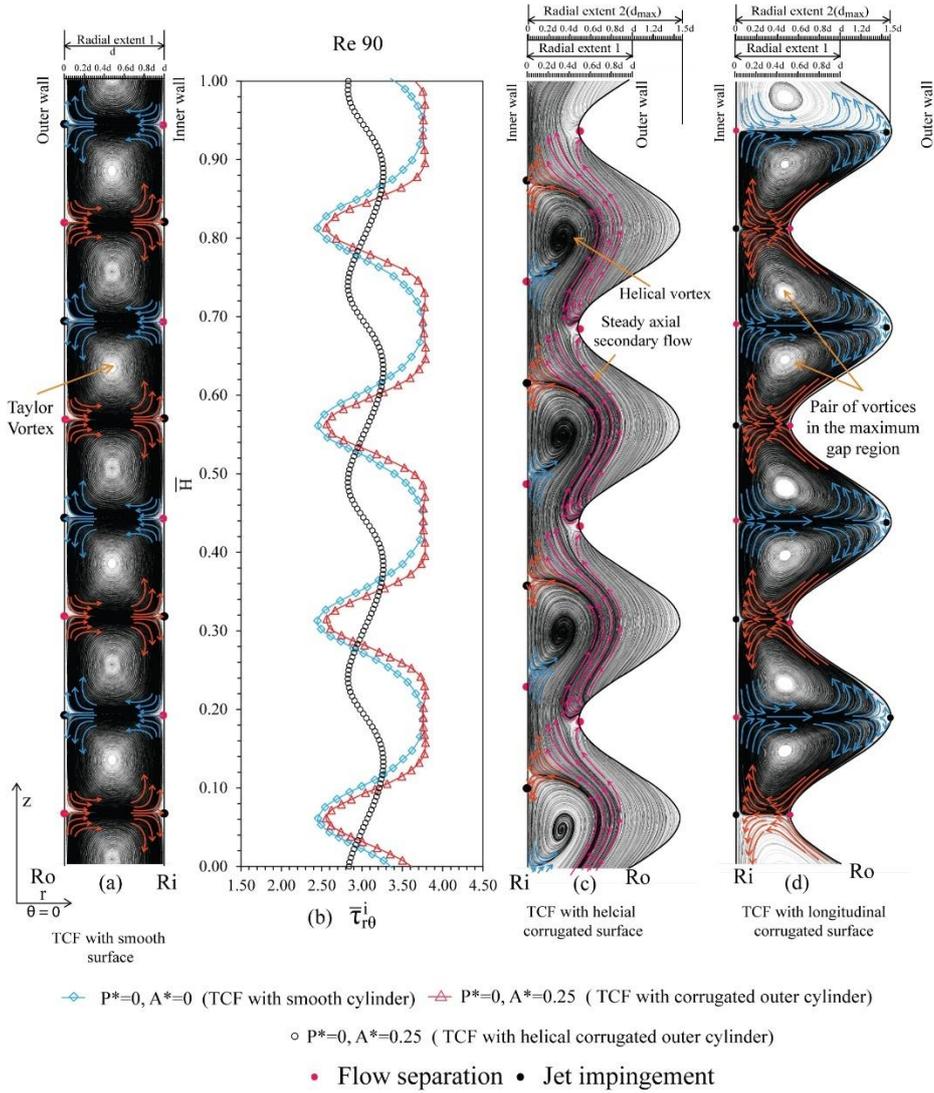

P*=0, A*=0 (TCF with smooth cylinder)  P*=0, A*=0.25 ( TCF with corrugated outer cylinder)

P*=0, A*=0.25 ( TCF with helical corrugated outer cylinder)

• Flow separation  • Jet impingement

*Figure 13: (a) Streamlines patterns associated with flow structure for TCF with the smooth wall(TCF_{Smooth}) at Re = 90 (reproduced from [Razzak, M. A., Khoo, B.C. & Lua, K.B., "Numerical study on wide gap Taylor Couette flow with flow transition". Physics of Fluids, 31(11) (2019)], with the permission of AIP Publishing )(b) The variation of shear stress along the axial direction for TCF with the smooth wall (TCF_{Smooth}), TCF with the longitudinal corrugated surface (TCF_{Longitudinal}) (A\* = 0.25) and helical corrugated surface (A\* = 0.25, P\*=1) at Re = 90 (c) Streamlines patterns associated with flow structure for TCF with longitudinal corrugated surface(TCF_{Longitudinal}) at Re = 90 (reproduced from [Razzak, M. A., Khoo, B.C. & Lua, K.B., "Numerical study of Taylor–Couette flow with longitudinal corrugated surface". Physics of Fluids 32, 053606 (2020)], with the permission of AIP Publishing ).(d) Streamlines patterns associated with flow structure for TCF with helical corrugated surface (TCF_{Helical}) at Re = 90.*

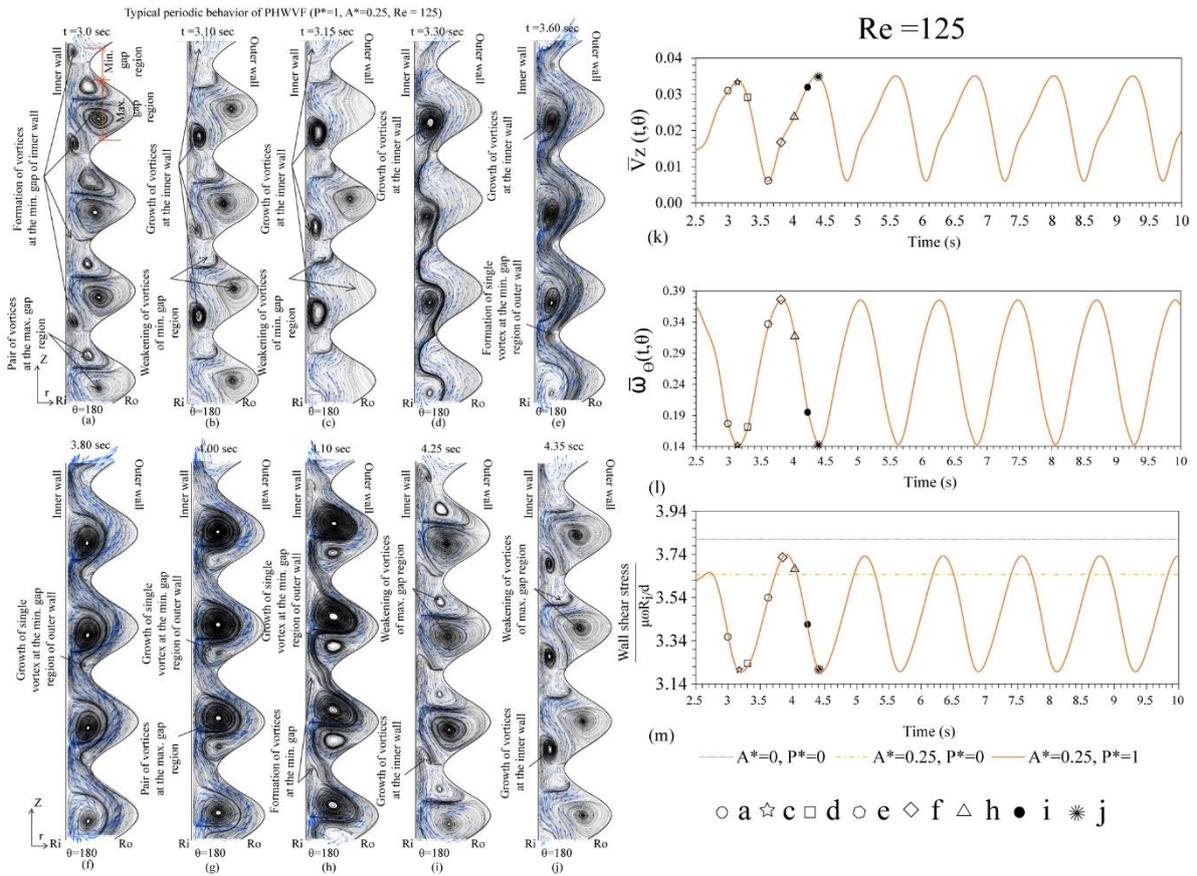

*Figure 14 : Typical transient behaviour of streamline pattern associated with flow structure at θ=180 for a single period of periodic helical wavy vortex flow (PHWVF) at Re = 125 (a)-(c) the emergence of a single vortex at the inner wall, its magnification and weakening and disappearance of pair of vortices at the maximum gap region (d) the replacement of pair of vortices of the maximum gap region by magnified newly formed single vortex (e) formation of another single vortex at the minimum gap region of the outer wall (f)-(g) shift of newly formed second single vortex along the axial direction and stays at the maximum gap region with the previously formed single vortex from the inner wall as pair of vortices (h)-(i) formation of another vortex at the inner wall and growth with time (k) transient variation of $\bar{V}_z(t,\theta)$ (l) transient variation of $\bar{\omega}_\theta(t,\theta)$ (m) transient variation wall shear stress at inner wall.*

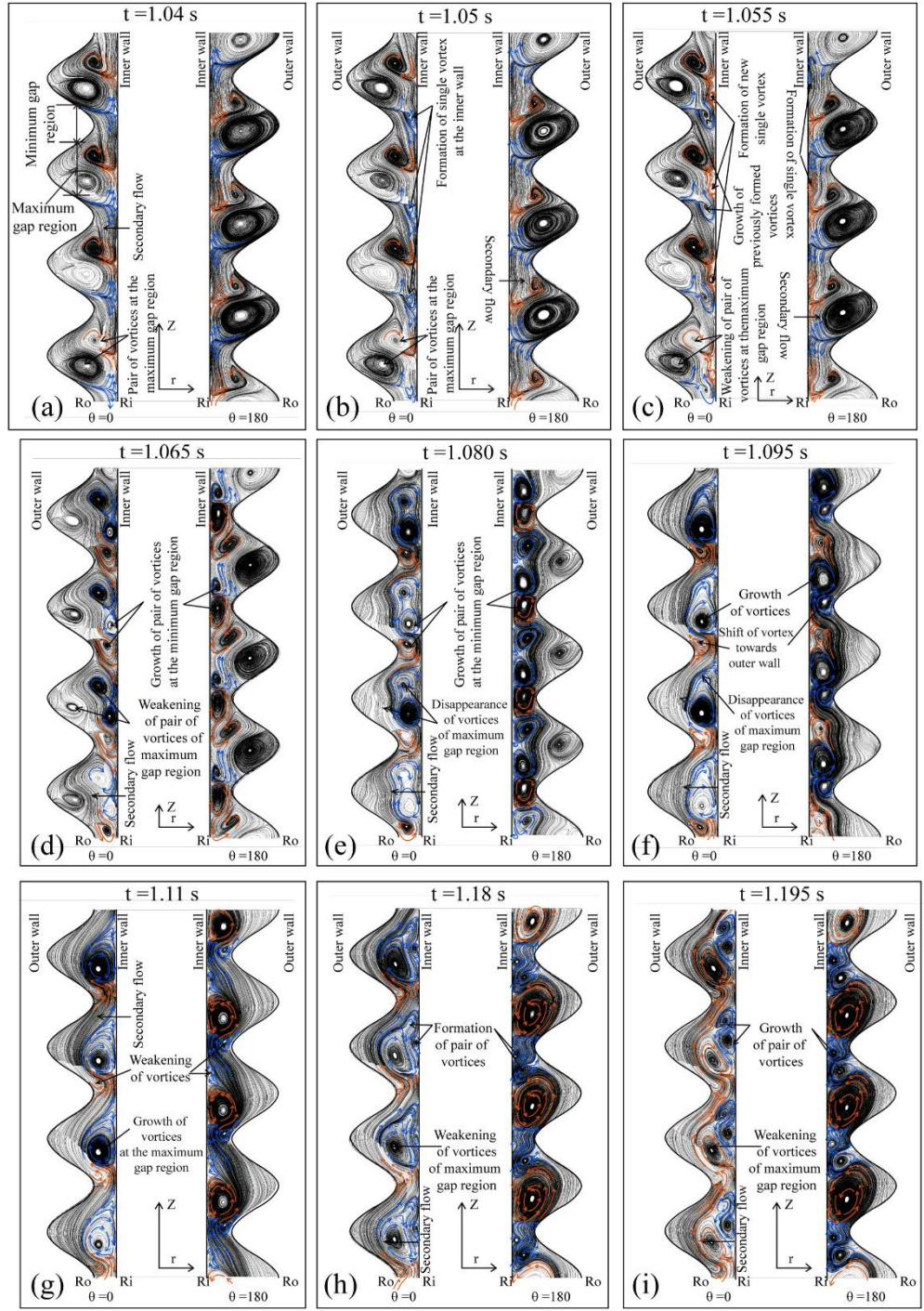

Formation of Non-periodic helical wavy vortex flow (NPHWVF)
P*=1, A*=0.25, Re=175

*Figure 15 : Typical transient evolution of instantaneous streamline patterns associated with the flow structure in the r-Z plane of θ=0 and 180 during the formation and growth of non-periodic helical wavy vortex flow (NPHWVF) for P* =1, A* = 0.25 and Re =175(a)-(b) formation of a single vortex at the inner wall of the minimum gap region (c) formation of another single vortex at the inner wall near to the maximum gap region (d)-(e) magnification of newly formed pair of vortices and their influence on weakening of pair of vortices of maximum gap region (f)-(g) further magnification of second vortex and taking over the place of the pair of vortices of the maximum gap region (h) disappearance of first vortex at the minimum gap region (i) formation of 3rd vortex at the minimum gap region of outer wall.*

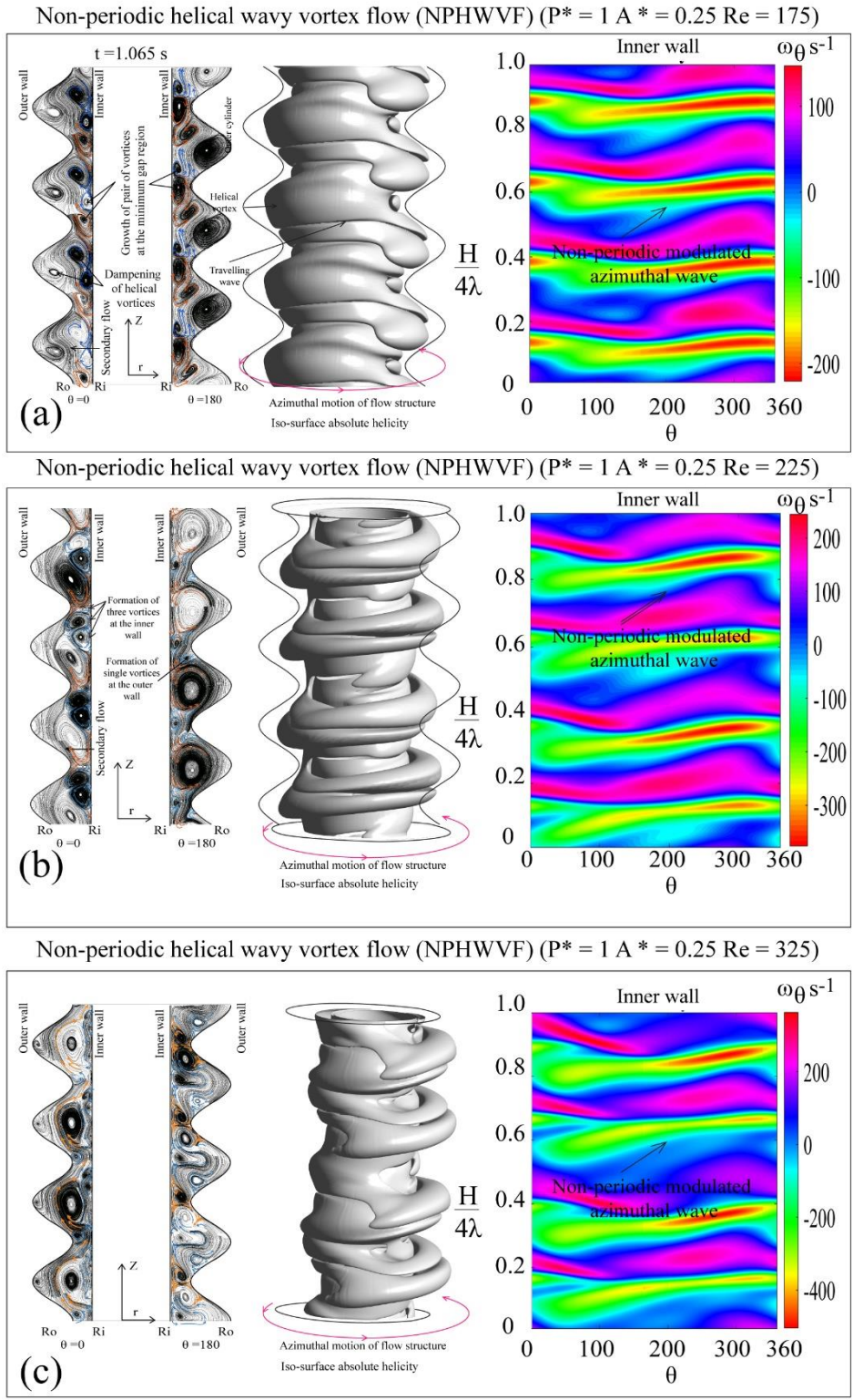

*Figure 16 : The typical variation of stream line pattern associated with flow structure in the r-Z plane of θ=0 and 180, absolute helicity and ω<sub>θ</sub> contour at the inner wall for non-periodic helical wavy vortex flow (NPHWVF) with Re for P\* =1 A\*= 0.25 (a) typical NPHWVF flow at Re = 175 and t=3 s (b)  typical NPHWVF flow at Re = 225 at t=3 s(c) typical NPHWVF flow at Re = 325 and t=3s.*